\documentclass[twocolumn]{aastex7}
\usepackage{amsmath,amstext}
\usepackage{graphicx} % Required for inserting images

\usepackage{xcolor}
\usepackage{natbib}
\graphicspath{{./figures/}}

\usepackage[shortlabels]{enumitem}

\defcitealias{Yao2023}{Y23}

\begin{document}

\title{Tidal disruption event rates across cosmic time: forecasts for LSST, Roman, and JWST and their constraints on the supermassive black hole mass function}
\shortauthors{Karmen et al.}
\shorttitle{High-z TDE rates}

% \suppressAffiliations 
\correspondingauthor{Mitchell Karmen}

\author[0000-0003-2495-8670]{Mitchell Karmen}
\affiliation{Department of Physics and Astronomy, Johns Hopkins University, 3400 N. Charles Street, Baltimore, MD 21218, USA}
\email{mkarmen1@jhu.edu}

\author[0000-0003-3703-5154]{Suvi Gezari}
\affiliation{Department of Astronomy, University of Maryland, College Park, MD 20742-2421}
\email{suvi@umd.edu}

\author[0000-0002-5222-5717]{Colin Norman}
\affiliation{Department of Physics and Astronomy, Johns Hopkins University, 3400 N. Charles Street, Baltimore, MD 21218, USA}
\email{cnorman3@jhu.edu}

\author[0000-0002-5063-0751]{Muryel Guolo}
\affiliation{Department of Physics and Astronomy, Johns Hopkins University, 3400 N. Charles Street, Baltimore, MD 21218, USA}
\email{mguolop1@jhu.edu}

\begin{abstract}
    Measuring the mass distribution of supermassive black holes (SMBHs) over cosmic time remains particularly challenging for the low-mass ($M_{\bullet}\lesssim10^8~M_\odot$) population at $z>1$. This population is also the most sensitive to SMBH seeding and early growth models. In this work, we construct a semiempirical model for the redshift evolution of the tidal disruption event (TDE) rate under multiple SMBH mass function prescriptions, and show that the observed redshift-dependent rate of TDEs is very sensitive to the SMBH mass function and its evolution with redshift. We further incorporate galaxy-scale processes that evolve with redshift-- namely, increasing galaxy nuclear stellar densities, enhanced galaxy-galaxy merger rates, dust obscuration, and a possible top-heavy initial mass function at early cosmic times-- and quantify their combined impact on the TDE rate. We find that including these effects generally results in a volumetric TDE rate that increases with redshift until a maximum near cosmic noon, before declining at higher redshift, where SMBHs that can disrupt stars become increasingly scarce. We forecast TDE rates in the Rubin Legacy Survey of Space and Time (LSST) and the Roman High Latitude Time Domain Survey, alongside expectations for serendipitous TDE rates in the JWST COSMOS-Web survey. Finally, we provide a methodology for using a flux-limited survey of TDEs in LSST to directly constrain the redshift evolution of the SMBH mass function.

\end{abstract}

\section{Introduction}

The origin of the supermassive black holes (SMBHs) that are believed to exist in the centers of most galaxies remains an enigma \citep{Volonteri2021, Rees1978}. The primary constraint on the seeding and growth of SMBHs comes from observations of massive quasars at high redshifts \citep{Inayoshi2020}, first in surveys such as the Sloan Digital Sky Survey (SDSS) \citep{Fan2001} and later in very deep ground-based surveys such as Subaru SHELLQS \citep{Matsuoka2018}. The latest studies performed with JWST have opened the door to the lowest-luminosity quasars, bringing the SMBH masses probed down to $M_{\bullet}\sim 10^6~M_{\odot}$ \citep{Maiolino2024Jades, Geris2026, Harikane2023}, but few sources at these low masses have been characterized so far across a broad range of redshifts.

Tidal disruption events (TDEs) are phenomena unique to these lower-mass SMBHs. A TDE occurs when a star passes near enough to a SMBH that the tidal forces from the SMBH overcome the star's self-gravity, shredding it apart. Main-sequence stars are visibly disrupted by nonrotating SMBHs with masses below the Hills mass \citep[$M_{\bullet}\lesssim 10^8\,M_\odot$,][]{Hills1975}, where the tidal disruption radius is beyond the Schwarzschild radius (i.e., the light from the disruption event escapes the SMBH). This upper limit is well observed in the mass distributions of SMBHs in TDEs \citep{vanVelzen2018, Yao2023, Mummery2024} with a small number of SMBHs beyond $10^8 M_\odot$ either due to the spin of the SMBH \citep{Kesden2012, Leloudas2016} or the disruption of a massive star \citep{Henkly2025}. It has been shown from first principles that the per-galaxy rates of TDEs are also sensitive to the stellar mass function of a galaxy \citep{Magorrian1999} and the volumetric rate of TDEs is sensitive to the volumetric SMBH mass function \citep{Stone2016, Kochanek2016}. To date, hundreds of TDEs have been detected \citep{Gezari2021, Yao2023, Mummery2024}, primarily by wide-field optical surveys \citep[e.g. ZTF,][]{Bellm2019}, but almost exclusively to very low redshifts $\leq 0.5$ \citep[except for rare jetted TDEs which have been observed to $z=1.2$, e.g.][]{Andreoni2022}. Within this small redshift volume and with the current TDE sample size, it is not possible to distinguish between the redshift evolution of different SMBH mass function models (as is shown later in this work explicitly). However, given upcoming wide-field surveys such as the Rubin Legacy Survey of Space and Time (LSST) and the Roman High Latitude Time Domain Survey (HLTDS) \citep{Ivezic2019, Rose2021} as well as deep observations from JWST surveys, TDEs will soon be discovered across a much broader redshift range.

% In this work, we predict the redshift-dependent TDE rate given a redshift-dependent SMBH mass function (hereafter BHMF). We aim to calculate an observation-driven rate, rather than a first-principles calculation, by combining the local observed TDE rate with the observed evolution of galaxy properties over cosmic time. We apply both semiempirical and simulated BHMF models, and observe a significant difference in their predicted TDE rates. We use this to forecast the observed TDE rate in LSST and Roman as a function of redshift, and show that their observed redshift distributions will place strong constraints on the low-mass end of the SMBH mass function. In addition, we take into account the evolution of galaxy properties which has been observed by early JWST observations: namely the evolution of stellar nuclear density, galaxy mergers, dust extinction/obscuration, and the IMF as a function of redshift. Each of these factors can affect the TDE rate depending on how it changes with redshift. We include these in our rate models and show that they have a major impact on the TDE rate, but are subdominant to the BHMF at $z>2$. 

In this work, we predict the redshift-dependent TDE rate given a redshift-dependent SMBH mass function (hereafter BHMF). We aim to calculate an observation-driven rate, rather than a first-principles calculation, by combining the local observed TDE rate with the observed evolution of galaxy properties over cosmic time. In \S\ref{sec:rate_calculation}, we outline our calculation of a redshift-dependent TDE rate. In \S\ref{sec:forecasts} we predict TDE rates for LSST, Roman, and JWST surveys. In \S\ref{sec:discussion} we discuss the implications of our redshift-dependent rates, and propose a methodology for constraining the BHMF using the LSST TDE sample. In \S\ref{sec:conclusion} we present our conclusions and perspectives on upcoming TDE discoveries. We adopt a \citet{Planck2020} cosmology: $H_0 = 67.4\ \mathrm{km\,s^{-1}\,Mpc^{-1}}$,  $\Omega_m = 0.315$, $\Omega_\Lambda = 0.685$. We treat the shape of a TDE spectrum as invariant with redshift and do not consider strongly lensed nor Population III TDEs. We assume a Hills mass of $10^8M_\odot$ and neglect explicit dependencies on black hole spin or AGN activity. Throughout, we use $\Gamma_{\rm gal}$ for per-galaxy TDE rates (yr$^{-1}$), $\Gamma_{\rm vol}$ for volumetric rates (Mpc$^{-3}$~yr$^{-1}$), and $\dot{N}$ for total observable counts (yr$^{-1}$).

\section{A semiempirical rate evolution}
\label{sec:rate_calculation}
Our goal in this section is to construct an observation-driven prediction for the redshift evolution of the TDE rate that can be directly compared to upcoming wide-field and deep surveys. Rather than deriving the TDE rate from first principles of stellar dynamics, we do a semiempirical calculation, beginning with the local, observed TDE rate and then modeling its evolution using empirically motivated redshift-dependent corrections.

We normalize the TDE rate to the local empirical measurement from ZTF (\cite{Yao2023} hereafter \citetalias{Yao2023}) and incorporate the effects of (i) the evolving supermassive black hole mass function (BHMF), (ii) redshift-dependent galaxy properties that modify stellar disruption rates, and (iii) observational selection effects given various surveys. We represent the local \citepalias{Yao2023} TDE rate with the double power-law fit to the rest-frame $g$-band luminosity function \citepalias[][their Equation (15)]{Yao2023}, $\phi_{L}$, in units of $\rm{Mpc}^{-3}yr^{-1}dex^{-1}$, and assume its shape does not change with redshift. We apply redshift-dependent dimensionless factors for dust obscuration $\mathcal{O}(z)$, BHMF evolution $N_{\rm BH}(z)$, and host galaxy effects $\mathcal{F}(z)$ to estimate a cumulative TDE rate $\dot{N}_{\rm TDE}$ in $yr^{-1}$  as follows:

    \begin{equation}
    \label{eq:tde_rate}
      \dot{N}_{\rm TDE} = \int_{0}^{z_{Ly}(\lambda)} \epsilon(z) \mathcal{F}(z) N_{\rm BH}(z) \dot{N}_0(z, \lambda) \mathcal{O}(z) dz
    \end{equation}

where $z_{Ly}$ is the redshift at which Lyman alpha absorption obscures all light that would be detected in a given filter at central wavelength $\lambda$. $\dot{N}_0(z)$ is the cumulative local empirical rate in \citetalias{Yao2023} per redshift bin, calculated as the local volumetric luminosity function integrated within the sensitivity of a given survey. This is parameterized in the $g$-band as: 

\begin{equation}
    \dot{N}_0(z, g\textrm{-band}) =  \frac{A_{\rm survey} }{ 4\pi} \frac{1}{1+z} \frac{dV_C}{dz} \int_{L_{\rm min}}^{L_{\rm max}} \phi_{L}(L_g) d \log {L_{g}}
    \label{eq:baseline_rate}
\end{equation}

\noindent where $\phi_L(L_g)$ is the $g$-band luminosity function (with units of $\mathrm{Mpc^{-3}\,yr^{-1}\,dex^{-1}}$), $\frac{dV_C}{dz}$ is the comoving volume in a given redshift bin, $L_{\rm min}$ is the luminosity limit of a given survey in the $g$-band at that distance $z$, and $L_{\rm max}=10^{44.7}~\rm{erg}~\rm{s}^{-1}$ is the luminosity of the brightest TDE observed in the \citetalias{Yao2023} sample. The factor of $1+z$ in the denominator is to account for time dilation. We account for the total area of the survey by multiplying this spherical volumetric rate by $\frac{A_{\rm survey} }{ 4\pi~\rm{steradians}}$. This luminosity function is k-corrected to other (UV/optical/NIR) filters by assuming the TDE emission is a blackbody, and adjusting to the respective filter's sensitivity and magnitude limit. We assume the temperature distribution from \citetalias{Yao2023} by drawing 300 Monte Carlo samples directly from the discrete blackbody temperatures of the ZTF sample at each redshift $z$ in order to uniformly represent the observed TDE temperatures. We account for filter sensitivity curves using synthetic photometry performed in the \texttt{sncosmo} Python package \citep{barbary_2025_15019859}. The $g$-band rate integrates to a local total volumetric rate of $3.1^{+0.6}_{-1.0} \times 10^{-7}~\rm{TDEs}~\rm{Mpc}^{-3}~\rm{year}^{-1}$. Note that the empirical rate is based on the TDE luminosity function rather than the SMBH mass function. This approach assumes that the shape of the luminosity function is constant with redshift. We discuss the merits and limitations of this assumption in Section \ref{sec:limitations}.  We propagate the uncertainties on the integrated TDE rate as uncertainties in the normalization of the luminosity function.

$\epsilon$ is defined as the ``efficiency'' of the survey. It is the fraction of events occurring in a given year that are seen by a survey: 

% \begin{linenomath*}
\begin{equation}
\epsilon(z) =
\left\{
\begin{array}{ll}
1, & \text{Time domain survey},\\[2pt]
\frac{T_{\textrm{visible}}(1+z)}{365~\mathrm{days}}, & \text{Single epoch survey}.
\end{array}
\right.
\end{equation}
% \end{linenomath*}

\noindent and is used to consider surveys with only a single epoch of observation. 

$T_{\rm visible}$ is the total rest-frame time for which the TDE satisfies $m_{\rm TDE} < m_{\rm limit}$ in the deepest filter, and the factor of $(1+z)$ accounts for cosmological time dilation. Therefore $\epsilon$ is dimensionless and can be interpreted as the probability that a TDE occurring at a random time is observable by a single-epoch survey.

To compute $T_{\rm visible}$, we model TDE light curves using the parameterization of \citet{vanVelzen2021}: 

    \begin{align}
    L_\nu(t) &= L_{\nu_0\, \rm peak}~\frac{B_\nu(T_0)}{B_{\nu_0}(T_0)} \nonumber \\
      &\times \begin{cases} e^{-(t-t_{\rm peak})^2/2\sigma^2} & t\leq t_{\rm peak} \\ 
      e^{-(t-t_{\rm peak})/\tau} & t>t_{\rm peak}\\
      \end{cases}
      \label{eq:lightcurve} 
    \end{align} 

with the luminosity in the $g$-band $42.68 < \log(L_g / \textrm{erg s}^{-1}) < 44.68$, rise time $0.4 < \log(\sigma / \textrm{days}) < 1.3$, and exponential decay time $1.2 < \log(\tau / \textrm{days}) < 2.3$, as seen in ZTF. $B_\nu(T_0)$ is the luminosity in the target filter that we model, and $B_{\nu_0}(T_0)$ is the luminosity in the reference filter, the $g$-band (where $\nu_0$ is the central frequency of the $g$-band), where we set the TDE luminosity. We simulate TDEs with all of the observed values of $\tau$ and $\sigma$ from the initial ZTF sample of spectroscopically confirmed TDEs \citep{Hammerstein2021}. From these simulated light curves, we compute $T_{\rm visible}$ with uncertainties given by the observed scatter in light-curve parameters, and interpolate $\epsilon = \epsilon(\Delta m)$, where $\Delta m \equiv m_{\rm peak} - m_{\rm limit}$, the amount by which the peak magnitude of the TDE exceeds the survey $5\sigma$ detection limit . This efficiency correction is applied to all non–time-domain surveys in Equation~\ref{eq:tde_rate}. For time-domain surveys, we assume sufficient cadence to detect TDEs independent of their phase.

We now turn to the physical evolution of the TDE rate. The dimensionless factor $\mathcal{F}(z)$ is the enhancement in TDE rate due to redshift-dependent galaxy properties. We define it as

\begin{equation}
    \label{eq:galaxy-effects}
    \mathcal{F}(z) \equiv \mathcal{M}(z)\mathcal{I}(z)\mathcal{D}(z)
\end{equation}

This is made up of three main properties: \begin{enumerate}
    \item $\mathcal{M}(z)$: the evolving galaxy merger rate
    \item $\mathcal{I}(z)$: the evolving initial mass function (IMF)
    \item $\mathcal{D}(z)$: the evolving nuclear stellar densities
\end{enumerate}

All three factors are normalized to unity at $z=0$.  In the following section, we discuss how these are estimated. Given a range of plausible models for each redshift-dependent rate modification, we propagate all uncertainties using Monte Carlo sampling. We randomly draw parameters for each realization from their respective probability distribution, and compute a mean TDE rate along with uncertainty intervals at each redshift step.

\subsection{Black hole mass function evolution}
\label{sec:BHMF}

The TDE rate is very sensitive to the SMBH mass function \citep{Stone2016}. More precisely, the volumetric TDE rate is sensitive to the volume density of central SMBHs that can visibly disrupt stars: those with $10^5 \lesssim M_\bullet/M_\odot\lesssim10^8$ (for non-spinning SMBHs and lower main-sequence stars). We choose the lower limit of $10^5~M_\odot$ based on the lowest-mass SMBH observed in \citepalias{Yao2023}. This mass function is well-understood for higher-mass SMBHs which host AGN up to high redshifts \citep{Inayoshi2020}. However, while models of the low-mass end of the BHMF tend to agree at $z=0$, by $z=2$ the predictions made by semiempirical models \citep{Merloni2008, Shankar2009}, analytical models \citep{Shen2009, Volonteri2010}, and cosmological simulations \citep{Hopkins2008} diverge \citep{Kelly2012}. Broad-line AGN are used to trace the BHMF as a function of redshift, but they primarily constrain the highest SMBH masses and may contain systematic uncertainties at higher redshifts \citep{Bertemes2025}. Recent work with JWST has extended these measurements to SMBHs with $10^7 \lesssim M_\bullet / M_\odot \lesssim 10^8$ in a single bin over the broad redshift range $3.5 < z < 6.0$, but with poor resolution in both redshift and mass space \citep{Taylor2025}. Additionally, recent studies \citep{Harikane2023, Maiolino2024, Taylor2025} suggest that SMBHs may be ``overmassive'' with respect to their cosmic epoch and host galaxies at high redshifts, but this population is not yet well-sampled \citep{Geris2026}.

Given the lack of ground truth for evolving the low-mass end of the SMBH mass function with redshift, we consider two contrasting models: the \texttt{Illustris} cosmological simulations \citep{Genel2014, Sijacki2015} and a semiempirical growth model \citep{Shankar2009} calibrated to observed AGN \citep{Hopkins2007}. Note that the \citet{Shankar2009} model is the same model used in the \citet{Kochanek2016} prediction of the TDE rate as a function of redshift, and indeed our results agree with theirs given the same inputs (when normalizing \citet{Kochanek2016} calculation to the more recent \citetalias{Yao2023} rather than to \citet{vanVelzen2014}). Also note that different semiempirical models of the BHMF diverge beyond $z>2$, especially at the low-mass ($M_\bullet\lesssim10^{8.5}$) end (compare, e.g. \citet{Shankar2009}, \citet{Kelly2010}, and \citet{Schulze2010}). We select these two models not because we believe them to be the most correct, but because they contrast strongly, meaning the true TDE rate should lie somewhere between the predictions of these two models\footnote{The black hole mass functions used in this work were also selected in part because of their public data accessibility. The authors invite any other redshift-dependent BHMF model or simulation which would like to be included in this comparison to reach out to the corresponding author.}.

The \citet{Shankar2009} model constructs a self-consistent model of SMBH growth which is directly calibrated to the bolometric luminosity function of AGN, which is estimated through a combination of X-ray and optical measurements \citep{Hopkins2007}. They use a continuity equation for the SMBH number density which applies the \citet{Soltan1982} argument to connect integrated AGN luminosity to black hole mass growth under the assumption that accretion dominates the assembly of SMBHs. This argument is powerful for testing a solely accretion-driven growth of SMBHs. However, because of lack of observational constraints on low-mass $M_{\bullet}\lesssim 10^7~M_{\odot}$ SMBHs at $z>1$, the evolution of these SMBHs at higher redshifts, and specifically the ``downsizing'' effect around cosmic noon, is largely determined by prescriptions for radiative efficiency and Eddington ratios. 

The \texttt{Illustris} simulation \citep{Genel2014} seeds SMBHs based on dark matter halo mass, inserting a black hole of mass $M_{\bullet}=1.42\times 10^5~M_{\odot}$ in halos with masses $M_{\rm halo}\geq 7.10 \times 10^{10}~M_{\odot}$. These black holes then accrete following the Bondi--Hoyle--Lyttleton \citep{Bondi1952} model up to a limit of the Eddington rate. \texttt{Illustris} then follows a feedback model to regulate the density of the ISM surrounding the black hole. This simulation therefore considers a range of accretion rates, as is observed in the local Universe \citep{Heckman2004}, and includes mergers of SMBHs. However, it too makes a range of assumptions-- such as fixing SMBHs to the centers of halos, ignoring relative BH velocities in the merger criterion, and artificially boosting accretion rates to compensate for under-estimated interstellar medium (ISM) densities surrounding the SMBHs. 

\begin{figure*}
    \centering
    \includegraphics[width=0.99\textwidth]{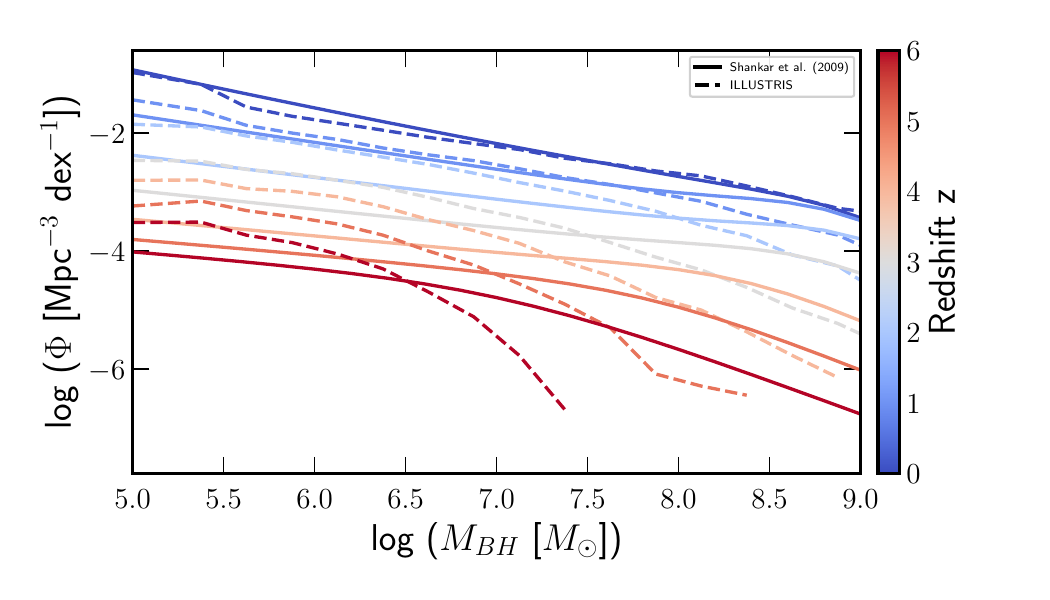}
    \caption{The two SMBH mass functions used in this work at $z\sim0$ through $z=6$. The solid line is the semiempirical model from \citet{Shankar2009} while the dashed line is the \texttt{Illustris} simulation \citep{Genel2014}. }
    \label{fig:BHMF_local}
\end{figure*}

For each model, we obtain a density $\phi_{\rm BH}(M,z)$ in $\rm{Mpc}^{-3} \rm{dex}^{-1}$, which is the number density of SMBHs within a given logarithmic mass bin at a given redshift. These two BHMFs are shown evolving with redshift in Figure \ref{fig:BHMF_local}. At higher redshifts, the range of black hole masses for which the BHMF is modeled is limited. At these redshifts, we truncate the BHMF where model predictions end. Generally speaking, the simulated BHMF is more ``bottom-heavy'' than the semiempirical BHMF at higher redshifts. 

At a given redshift $z$, we calculate the volume density of SMBHs that can disrupt main-sequence stars $10^5 \lesssim M_\bullet / M_\odot \lesssim 10^8$, which is bounded from below by the physical sizes of main-sequence stars \citep{Gezari2021} as well as the observational lower limit of central SMBHs seen to cause TDEs, and from above by the Hills mass \citep{Hills1975}. That is: \begin{equation}
\label{eq:n}
    n_{\rm BH}(z) = \int_{10^5 M_{\odot}}^{10^8 M_{\odot}} \phi_{\rm BH}(M,z)d\log M
\end{equation} 

We scale the empirical local TDE rate from ZTF \citepalias{Yao2023} by the relative volume density of these SMBHs at a given redshift by the local density: \begin{equation}
    N_{\rm BH}(z) = \frac{n_{\rm BH}(z)}{n_{\rm BH}(z=0)}
\end{equation}

Which leads to the $N_{\rm BH}(z)$ term that is introduced in Equation \ref{eq:galaxy-effects}.

\subsection{Dust obscuration}
\label{sec:obscuration}

\begin{figure}
    \centering
    \includegraphics[width=\linewidth]{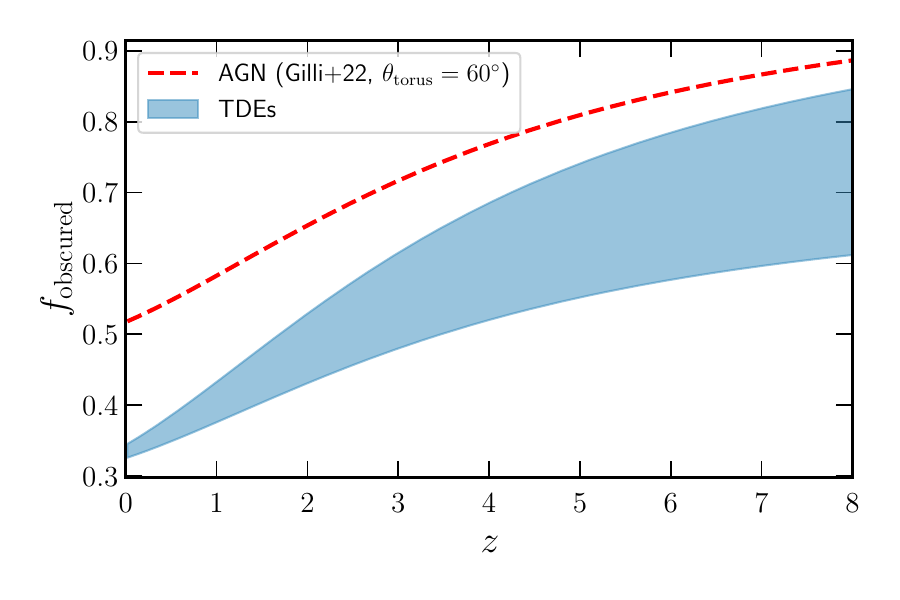}
    \caption{Evolution of the assumed TDE obscuration fraction as a function of redshift, shown by the blue shaded region. The redshift evolution of the obscuration fraction for typical AGN, as modeled in \citet{Gilli2022} is shown in red. The TDE obscuration fraction is lower at low redshifts because of the lack of a dusty torus, but grows with ISM column density.}
    \label{fig:dust}
\end{figure}

The local \citetalias{Yao2023} TDE rate does not correct for obscured TDEs. Recent work has demonstrated that a population of optically obscured TDEs exists in the local universe, detected through mid-infrared dust emission. \cite{Masterson2024} used \textit{NeoWISE} data to identify a class of TDEs that lack optical counterparts but produce mid-IR flares inconsistent with supernovae or AGN variability. They are frequently found in uncommon host galaxies for TDEs- instead of the typical green valley, compact host; these IR-only TDEs occur in massive, star-forming galaxies. The IR TDEs have an estimated volumetric rate of $1.3 \times 10^{-7}~\rm{Mpc}^{-3}~\rm{yr}^{-1}$ which, when taken together with the optical TDE sample, would imply a local obscuration fraction of $f_{\rm obsc}\sim 30\%$. This estimation ignores the samples of X-ray detected TDEs \citep[e.g. ][]{Guolo2024}, but is useful for estimating observed rates of optically selected TDEs. To approximate the evolution of the TDE obscuration fraction as a function of redshift, we look to observations of AGN.

TDE obscuration and AGN obscuration in the local universe arise from predominantly different phenomena. While TDEs are likely obscured by their host galaxies' star formation and dusty ISM, AGN are believed to be obscured by a dusty torus that they generate \citep{Netzer2015} as well as their host ISM. AGN spectra do not change significantly as a function of redshift \citep{Shen2019}, implying that their structure and physics do not change at higher redshifts. However, the obscuration fraction of AGN is seen to increase with redshift, which suggests that this is due to changes to their host galaxies \citep{Gilli2022}; the increasing galaxy densities and dust masses as a function of redshift lead to a higher-column-density ISM, which in turn obscures a larger fraction of AGN. 

We use the redshift evolution of the AGN obscuration fraction as an empirical tracer of the evolving ISM column densities and dust content of massive galaxy hosts. Although optical TDEs at low redshift preferentially occur in different host galaxies than AGN, the existence of an obscured TDE population in massive, star-forming galaxies indicates that TDE environments span a wide range of ISM conditions. We thus assume that the redshift evolution of dust obscuration in TDE hosts generally follows the same trend as the redshift evolution of dust obscuration in AGN host galaxies.

We divide TDEs into two classes—``obscured'' and ``unobscured''—and parameterize the obscuration fraction with a logistic function to ensure that it smoothly transitions from a given local obscuration fraction to a saturated value at higher redshift:
\begin{equation}
f_{\mathrm{obsc}}(z)
=
f_{0}
+
\frac{f_{\max}-f_{0}}
{1 + \exp\!\left[-k\,\log\!\left(1+z\right)\right]}
\end{equation}

We set $f_0$ as the local TDE obscuration fraction of $f_{\mathrm{obsc}}(z=0) = 0.3$. The redshift-dependence parameter $k = 0.7$ is calibrated to AGN obscuration measurements \citep{Gilli2022}.We let $f_{\rm max}$ vary from the lowest obscuration fraction, $f_{\rm max}=0.8$, to the highest obscuration fraction, $f_{\rm max}=0.95$, predicted for AGN with a range of ISM column densities \citep{Gilli2022} to account for uncertainty in the high-redshift obscuration fraction. We then include this obscuration fraction in Equation \ref{eq:galaxy-effects} as the dimensionless scaling relative to the local universe: \begin{equation}
    \mathcal{O}(z) = \frac{f_{\mathrm{obsc}}(z)}{f_{\mathrm{obsc}}(z=0)}
\end{equation}

The resulting assumed obscuration fraction as a function of redshift is seen in Figure \ref{fig:dust}. We note that this approach is intentionally phenomenological, and neglects effects such as partial obscuration or wavelength-dependent attenuation. Considering this to be an order-of-magnitude estimate of the effect of obscuration, it ends up as a very minor factor in Equation \ref{eq:galaxy-effects} as seen in the following results.

\subsection{Nuclear stellar density}

\begin{figure*}
    \centering
    \includegraphics[width=\linewidth]{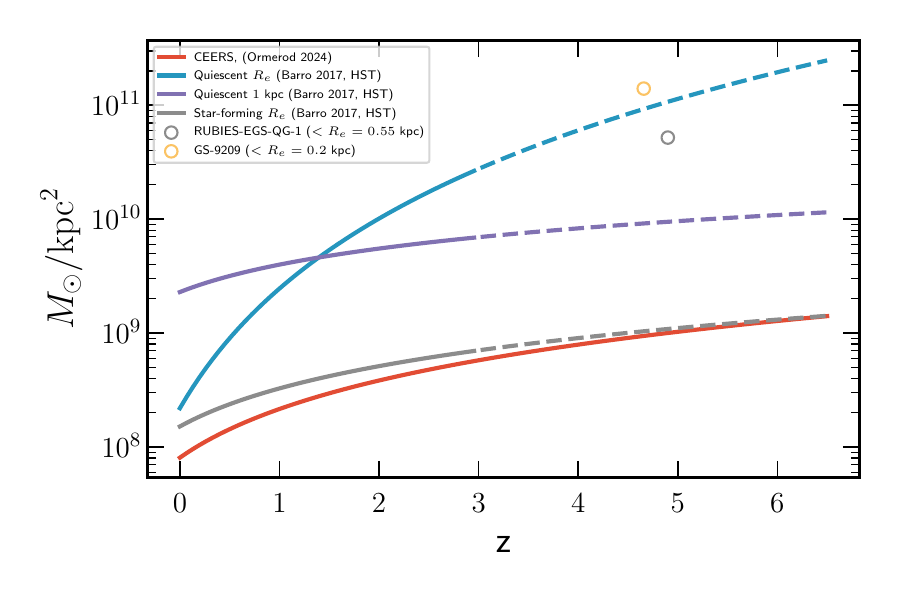}
    \caption{Evolution of galaxy central surface density as a function of redshift. Lines are solid where measured, and dashed where \cite{Barro2017} is extrapolated to higher redshifts. It can be seen that the extrapolation of the star-forming sample is a good prediction of what is later observed in the CEERS \citep{Ormerod2024, Finkelstein2023} data taken with JWST. The quiescent-galaxy extrapolation also predicts the two highest-redshift quiescent galaxies for which this density has been measured, \citep{deGraaff2025, Carnall2023}. We use the $1$~kpc extrapolation to predict the evolution of central galaxy density.} 
    \label{fig:density}
\end{figure*}

The rate of TDEs in a given galaxy depends directly on the density of stars within the SMBH's radius of influence, $r_{\rm inf}$. However, the stellar density within $r_{\rm inf}$ is extremely difficult to measure beyond very local galaxies; for a typical TDE host, $r_{\rm inf} \sim 5-50$~pc. In this section, we outline our methodology of evolving this density with redshift.

It is expected from theories of galaxy formation that early galaxies would be small and dense relative to their low-redshift counterparts \citep{White1978, Mo1998}. This is seen in cosmological simulations \citep{Dubois2021} and has been observationally confirmed by deep \textit{Hubble Space Telescope} imaging \citep{vanDerWel2014}. JWST has now revealed extraordinarily compact systems in the early universe \citep{Baggen2023, Guia2024}, indicating that dense stellar environments were not uncommon in the early Universe. These very dense systems may have more stars near their central SMBHs which can be disrupted, leading to an enhanced TDE rate at higher redshifts. 

As mentioned, the TDE rate in a given galaxy depends on the density within the radius of influence of the SMBH, $\rho(r_{\rm inf})$, rather than the global galaxy density. To calculate this dependence, \citet{Lightman1977} solved the averaged Fokker--Planck equation, effectively calculating the TDE rate for a population of stars around an SMBH. They find that stars with low enough angular momentum, $J < J_{\rm min}(E) \sim \sqrt{GMr_t}$ enter the ``loss cone'', and are therefore tidally disrupted. This divides TDEs into two regimes, one in which the loss cone is ``full'' and the flux of TDEs is limited by the orbital period over which the orbits reach $r<r_t$, and one in which the loss cone is ``empty'' and the TDE rate is limited by the angular momentum diffusion into the loss cone. 

In the full-loss-cone regime, the TDE rate, $\Gamma_{\rm gal}$, is simply the number of stars with orbits within the loss cone, divided by their orbital period. Therefore, assuming an isotropic distribution of orbits, the overall TDE rate is proportional to the number of stars within $r_{\rm inf}$ and thus scales linearly with the central density of the galaxy: $\Gamma_{\rm gal} \propto \rho(r_{\rm inf})$. In the empty-loss-cone case, the flux into the loss cone is the total number of stars near the critical radius divided by their two-body relaxation time-- the time it takes for them to scatter each other into low-angular-momentum orbits, $\Gamma_{\rm gal} \sim N / t_{\rm rel}$. Since $N \propto \rho(r_{\rm inf})$ and $t_{\rm rel} \propto \rho^{-1}$ \citep[][Equation 2]{Lightman1977}, the overall TDE flux scales as $\Gamma_{\rm gal} \propto \rho^2(r_{\rm inf})$. The overall TDE rate must scale at some intermediate point between, \begin{equation}
    \Gamma_{\rm gal} \propto \rho^\alpha(r_{\rm inf})
    \label{eq:density_ndot}
\end{equation}

\noindent where the exponent $\alpha$ is bounded by them: $1 \leq \alpha \leq 2$. It has been found that for classical bulges, the TDE rate leans toward the diffusion-limited regime implying $\alpha \sim 2$ \citep{Wang2004}. \citet{Stone2016} computed TDE rates for a large sample of nearby galaxies and showed that only for the lowest-mass and most weakly-concentrated nuclei is the full loss cone scenario more likely. More recent studies likewise find that the relative importance of full- and empty-loss-cone feeding can vary substantially across galaxy populations and black hole masses \citep{Chang2025, Hannah2025b}. These models consistently predict that denser galactic nuclei yield higher TDE rates, motivating the parameterization in Equation~\ref{eq:density_ndot}. Observationally, \citet{Graur2018} finds a near-linear relationship between the TDE rate and surface mass density. The surface mass density is measured within $1~$kpc, not $r_{\rm inf}$, so this cannot directly be compared to theoretical predictions; a reasonable galaxy surface density profile can render this scaling consistent with the mass density within $r_{\rm inf}$.

 Direct measurements of $\rho(r_{\rm inf})$ are unavailable for the high-redshift galaxy population considered in this work. We instead use resolved central surface-density measurements on larger scales as empirical tracers of the nuclear stellar environment. This approach is motivated by local observations showing that TDE hosts are preferentially green-valley, centrally concentrated galaxies with elevated S\'ersic indices relative to control samples \citep{Hammerstein2021}, and by high-resolution imaging which finds enhanced stellar mass surface densities on $\sim30$--$100$ pc scales in several nearby TDE hosts \citep{Graur2018, French2020}. Studies of nuclear star clusters indicate that parsec-scale nuclear densities correlate with host-galaxy mass and structural properties \citep{Neumayer2020,Pechetti2020,Hannah2025a}. No direct calibration currently exists between densities measured within $\sim1$ kpc and those at $r_{\rm inf}$, although these results suggest that galaxy-scale central densities retain some information about unresolved nuclear conditions. 

%However, by assuming a roughly constant density profile shape with redshift, we use the surface brightness density measurements at larger scales to trace this central density. 

The overall mass--radius evolution since $z=8$ has been measured using JWST within the CEERS field \citep{Ormerod2024}. For a fixed stellar mass, they parameterize: 
\begin{equation}
    r(z) \propto (1+z)^{-0.71}
\end{equation}

\noindent which implies $\rho(z) \propto (1+z)^{2.1}$ when averaged over galaxy effective radii. The effective radii of the galaxies in this study range from a typical $3$~kpc at $z=0$ to $\leq 1$~kpc at $z=7$. Studies have also focused on to the central regions of quiescent galaxies, which host most optically-selected TDEs. It is found that the central kiloparsec of quiescent galaxies up to $z\sim 3$ exhibits a more modest scaling in density of $\rho_{1 \rm kpc}(z) \propto (1+z)^{0.9}$ \citep{Damjanov2011, Barro2017}. Observations of very high redshift quiescent galaxies show high densities at smaller scales, such as RUBIES-EGS-QG-1 \citep{deGraaff2025} and GS-9209 \citep{Carnall2023}, which have densities $\rho = 7.1\times 10^{10}~M_{\odot}~ \rm{kpc}^{-3}$ within $0.55$~kpc and $\rho = 4.9 \times 10^{11}~M_{\odot}~ \rm{kpc}^{-3}$ within $0.2$~kpc respectively. \citet{Baggen2023} has found central densities as high as $\rho = 10^{12}~M_{\odot}~ \rm{kpc}^{-3}$ which are high compared to the local universe, but comparable to the densest compact quiescent galaxies (``red nuggets'') at cosmic noon \citep[e.g. ][]{Bezanson2009}. 

We compare the densities of the aforementioned high-redshift quiescent galaxies to the densities predicted by observed mass-scaling relations in Figure \ref{fig:density}. We find that, while the HST-derived \citep{Barro2017} density-redshift relations are only calculated up to $z=3$, extrapolating the relation for star-forming galaxies to $z=6$ yields broad agreement with observations from CEERS obtained with JWST \citep{Ormerod2024}. Furthermore, the extrapolation of density within effective radius $R_e$ for quiescent galaxies accurately reproduces the few high-redshift quiescent systems observed. Therefore, we assume that the evolution of the central density within 1~kpc can also be extrapolated to $z=6$. Therefore, we adopt the observed evolution of stellar density within the central $1$~kpc as a phenomenological proxy for the redshift evolution of nuclear stellar densities relevant to TDE production. This implies that while galaxies become increasingly compact with redshift, the resolved stellar densities of their inner regions would also increase modestly beyond $z\sim2$, consistent with current observations. We parameterize this trend as
\begin{equation}
\rho_{\rm central}\propto(1+z)^{0.9},
\end{equation}
while emphasizing that the mapping between this resolved quantity and the true density at $r_{\rm inf}$ remains uncertain.

%We therefore adopt $\rho_{\rm central}\propto (1+z)^{0.9}$ as our evolution of nuclear stellar density with redshift. We note that this is likely a conservative estimate but treat it as a proxy for the general evolution of galaxy centers.

Overall, assuming that $1 \leq \alpha \leq 2$ is reasonable in Equation \ref{eq:density_ndot}, we approximate the dependence of the TDE rate on the redshift evolution of galaxy density. This yields the dimensionless scaling term in Equation \ref{eq:galaxy-effects}: 

\begin{equation}
    \mathcal{D}(z) = (1+z)^{0.9 \alpha}
\end{equation}

For this parameter, we adopt a uniform prior for $1 \leq \alpha \leq 2$. While we adopt the conservative estimate for nuclear stellar density here, extreme stellar cluster densities have been observed in the early universe \citep{Vanzella2023, Adamo2024} which would result in extreme enhancements of the TDE rates in these systems \citep{Kritos2025}. We assume these systems are the minority and note that our scaling is likely a lower limit on the true TDE rate.

\subsection{Galaxy mergers}

The galaxy merger rate evolves strongly with redshift, peaking around cosmic noon ($z\sim2$). Observations suggest that TDEs preferentially occur in postmerger galaxies \citep{French2020, Wevers2024a}. This is seen through the overrepresentation of TDEs in rare poststarburst (PSB) galaxies \citep{Arcavi2014, French2016, LawSmith2017}, which make up $<1\%$ of the overall galaxy population. These galaxies are green and centrally concentrated, two key properties that are very common in TDE hosts \citep{Hammerstein2021}. It is seen that among PSB galaxies, TDEs are further found to preferentially occur in those containing extended emission-line regions \citep[EELRs, ][]{Wevers2024a} with an increase by a factor of 10. These EELRs have very narrow emission line velocity dispersions, which make it more likely that they originate from gas-rich mergers than AGN-driven outflows. These host galaxy preferences are also exhibited by quasi-periodic eruptions \citep{Wevers2024} which are now believed to originate from TDEs \citep{Nicholl2024, Chakraborty2025, Bykov2025}. Furthermore, the TDE rate shows signs of enhancement in interacting galaxies preceding a poststarburst phase \citep{Onori2025}.

It is hypothesized that this enhancement could be due to chaotic orbits induced by the merger \citep{Arcavi2014}, or recent bursts of star formation triggered by the merger \citep{StonevanVelzen2016} that populated the loss cone. This star formation could result from a large number of mechanisms, such as the formation of an SMBH binary \citep{Stone2011}, high stellar densities \citep{Stone2018}, or eccentric nuclear stellar disks \citep{Madigan2018}. Hydrodynamical cosmological simulations likewise show that TDE rates are enhanced after a merger \citep{Pfister2019, Pfister2021} by a factor of $1-2$ orders of magnitude. Recently, it has been suggested that eccentric Kozai-Lidov oscillations can also increase the TDE rate around binary SMBH systems \citep{Melchor2025}, which has been extended to a redshift-dependent rate. This calculation also predicts a TDE rate that increases toward $z\sim 2$.

As a proxy for the galaxy merger rate as a function of redshift, we adopt the observed evolution of major-merger rates derived from pair counts in deep surveys \citep{Ventou2017, Duncan2019}, and assume these to be representative of the galaxies that may host TDEs. \citet{Ventou2017} uses deep MUSE spectroscopy within the Hubble Ultra-Deep Field and Hubble Deep Field-South to identify galaxy close pairs. These pairs have low relative velocities and a high probability of merging within $T_{\rm pair} < 1$~Gyr. These rates are generally in agreement with one another, and with simulations \citep[e.g. HORIZON-AGN, \texttt{Illustris},][]{Kaviraj2017, Snyder2017}. 

In the aforementioned simulations, it is found that the dimensionless rate enhancement $E$ after a merger ranges from $10\leq E \leq 100$ for a timescale of $100 \leq t_{\rm enh} \leq 300$~Myr. We write the volumetric TDE rate without any enhancements from mergers to be $\Gamma_{\rm vol, 0}$ in $\rm{Mpc}^{-3} \rm{yr}^{-1}$ as: 

% \begin{equation}
% \Gamma_0 = n \langle \Gamma_{\rm gal} \rangle
% \end{equation}

% \noindent where $n$ is the number density of galaxies in $\rm{Mpc}^{-3}$ and $\Gamma_{\rm gal}$ is the rate of TDEs in a galaxy in $\rm{year}^{-1}$. Therefore, the rate of TDEs in an average galaxy after a merger can be written

% \begin{equation}
%     \Gamma_{\rm merger} = E  \Gamma_{0}
% \end{equation}

\begin{equation}
\Gamma_{\rm vol,0} = n \langle \Gamma_{\rm gal,0} \rangle ,
\end{equation}

\noindent where $n$ is the number density of galaxies in $\rm{Mpc}^{-3}$ and  $\langle \Gamma_{\rm gal,0} \rangle$ is the average per-galaxy TDE rate in  $\rm yr^{-1}$ in the absence of merger-driven enhancements. For a galaxy in the  enhanced post-merger phase, we write the per-galaxy TDE rate as

\begin{equation}
    \Gamma_{\rm gal,merger} = E \, \Gamma_{\rm gal,0},
\end{equation}
\noindent where $E$ is the dimensionless enhancement factor.

 The fraction of time that a galaxy spends in a state with an enhanced TDE rate is approximated as

\begin{equation}
f_{\rm enh}(z) = f_{\rm pair}(z) \left( \frac{t_{\rm enh}}{T_{\rm pair}} \right).
\end{equation}

Where $f_{\rm pair}$ is the instantaneous major close pair fraction at redshift $z$. Because $T_{\rm pair} \sim 0.3 - 0.7$ Gyr \citet{Lotz2011}, we set $\frac{t_{\rm enh}}{T_{\rm pair}} \approx 1$. We use the pair fraction redshift relationship from \citet{Ventou2017}:
\begin{equation}
    f_{\rm pair} (z) = 0.056 (1+z)^{5.910} e^{-1.814(1+z)}
\end{equation} 

% The merger-enhanced volumetric rate of TDEs is therefore %%%% 
 
% \begin{equation}
% \Gamma_{\rm enhanced}(z) = n \,\langle \Gamma_0 \rangle \left[1 + (E-1)\, f_{\rm enh}(z)\right].
% \end{equation}

% We normalize the enhancement such that the local ($z=0$) volumetric rate remains unchanged.
% \begin{equation}
% \Gamma_{\rm local} \equiv n \langle \Gamma_0 \rangle \left[ 1 + (E - 1) f_{\rm enh}(0) \right],
% \end{equation}
The merger-enhanced volumetric rate of TDEs is therefore

\begin{equation}
\Gamma_{\rm vol,enh}(z) =
n \langle \Gamma_{\rm gal,0} \rangle
\left[1 + (E-1)\, f_{\rm enh}(z)\right].
\end{equation}

We normalize the enhancement such that the local ($z=0$) volumetric rate remains unchanged:

\begin{equation}
\Gamma_{\rm vol, enh}(z=0) \equiv
n \langle \Gamma_{\rm gal,0} \rangle
\left[1 + (E - 1) f_{\rm enh}(z=0) \right].
\end{equation}

the redshift-dependent merger enhancement factor becomes
\begin{equation}
\mathcal{M}(z)
\equiv \frac{\Gamma_{\rm vol, enh}(z)}{\Gamma_{\rm vol,  enh}(z=0)}
=
\frac{
1 + (E - 1) f_{\rm pair}(z) \left( \frac{t}{T_{\rm pair}} \right)
}{
1 + (E - 1) f_{\rm pair}(0) \left( \frac{t}{T_{\rm pair}} \right)
}.
\end{equation}

As expected, the enhancement peaks at $z\sim2$, at the peak of merger history, and then declines slowly with redshift. When calculating uncertainties in this rate enhancement, we let $E$ vary uniformly across the full possible range $10 \leq E \leq 100$.

\subsection{Initial mass function evolution}

The TDE rate, alongside the SMBH mass function, also depends on the stellar mass function. In the local universe, it has been proposed that the TDE rate can be used as a probe of the stellar mass function of the host galaxy \citep{Stone2016, D'Orazio2019}. \citet{Stone2016} point out that, as opposed to a distribution of only $M = M_\odot$ stars, using a stellar mass function affects the angular momentum diffusion coefficient $\bar{\mu} \propto \langle M_{\star}^2 \rangle$.  Therefore, if the initial mass function changes as a function of cosmic time, it should have an impact on the TDE rate. Additionally, the Hills mass changes for more massive stars, which should increase the total rate of TDEs \citep{vanVelzen2018, D'Orazio2019}.

Observations of highly UV-luminous galaxies at early cosmic times \citep[e.g.][]{Finkelstein2024} hint at the possibility of a top-heavy IMF \citep{Inayoshi2022, Finkelstein2023, Trinca2024}. If this is the case, it could also enhance the TDE rate at high redshifts.

To quantify the effect of a redshift-dependent IMF, we parameterize the IMF as a single-slope power law: \begin{equation}
    \xi (M) \propto M^{-\beta}; M \in [M_{\rm min}, M_{\rm max}]
\end{equation}

to compute the mean-squared stellar mass that sets the angular momentum diffusion coefficient. We assume a fixed mass-to-light ratio for massive stars (which dominate the UV continuum) of $L \propto M^{3}$ \citep{Kippenhahn1994} and solve for the enhancement in massive stars necessary to explain the UV luminosities of observed high-redshift galaxies in \cite{Finkelstein2024}. We assume a Salpeter IMF \citep[$\beta = 2.35$,][]{Salpeter1955} in the local universe, and fit a linear decrease in $\beta$ until it reaches $\beta = 2.081$, the slope necessary to reproduce UV-bright galaxies at high-redshift ($z\sim8$) \citep{Finkelstein2024}. We find that the assumption of linearity of $\beta$ with redshift makes a negligible impact on the result. We use this IMF to scale $\langle M_{\star}^2 \rangle$ with redshift, which in turn scales $\bar \mu$ which is proportional to the TDE rate.  We get a final dimensionless scaling:

\begin{equation}
\mathcal{I}(z) = \frac{\langle M_\star^2\rangle_{\beta(z)}} {\langle M_\star^2\rangle_{\beta_0}}
\end{equation}

We note that the second moment of the mass function, $\langle M_*^2 \rangle$, may be dominated by compact remnants, particularly stellar-mass black holes, rather than by main-sequence stars in evolved nuclear star clusters. Because $\langle M^2 \rangle$ weights the highest-mass objects most strongly, even a modest population of stellar-mass black holes can dominate the relaxation process that feeds the loss cone \citep[see Section 2.3 of][]{Stone2016}. The abundance and mass spectrum of such remnants are expected to evolve with redshift due to metallicity-dependent stellar evolution. Our simplified parameterization of $\mathcal{I}(z)$ does not account for this.

\section{Rate forecasts for wide-field surveys}
\label{sec:forecasts}
\subsection{The volumetric TDE rate}

\begin{figure*}
    \centering
    \includegraphics[width=\textwidth]{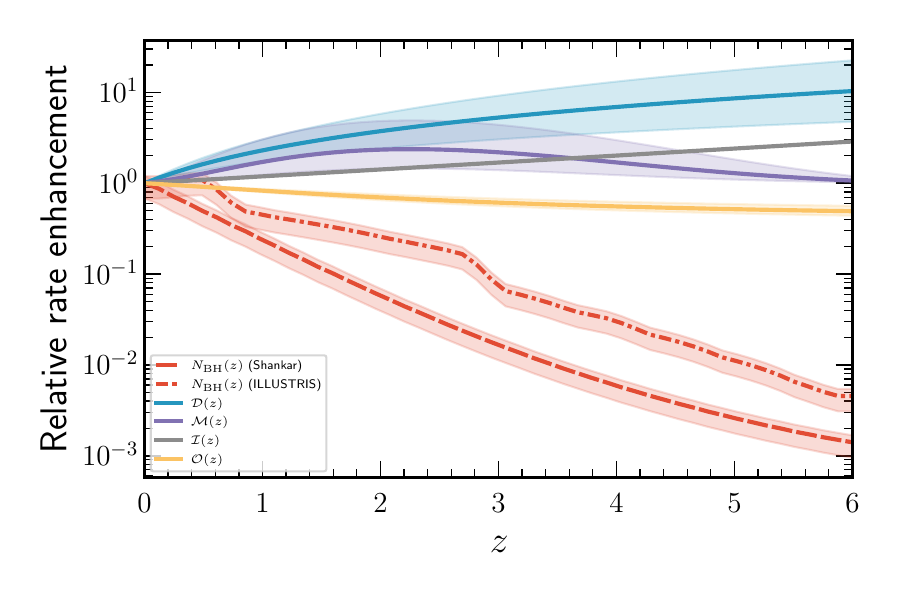}
    \caption{Scalings of all rate modifications as a function of redshift: the black hole mass function scaling ($N_{\rm BH}(z)$), the nuclear stellar density scaling $\mathcal{D}(z)$, the galaxy merger scaling $\mathcal{M}(z)$, the IMF scaling $\mathcal{I}(z)$, and the dust obscuration scaling $\mathcal{O}(z)$. On the left y-axis, a value of $1$ is equivalent to the local value. The shaded region represents the uncertainty propagated into the final rate estimate. A value of 1 for all scalings corresponds to the local TDE rate.}
    \label{fig:enhancements}
\end{figure*}

\begin{figure*}
    \centering
    \includegraphics[width=\linewidth]{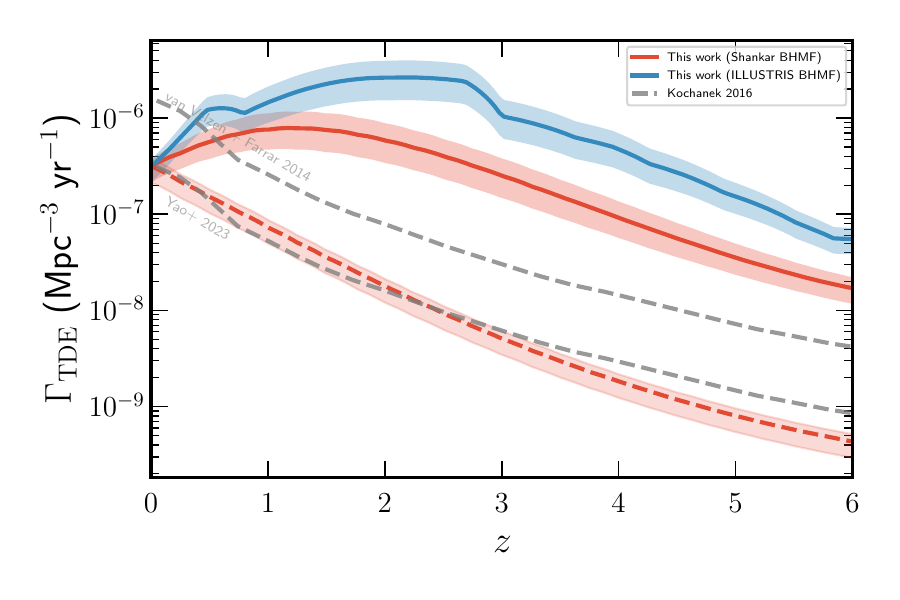}
    \caption{Volumetric rate of TDEs as a function of redshift after applying all redshift-dependent rate modifications. The red line uses the \cite{Shankar2009} BHMF model and the blue uses the \texttt{Illustris} simulation. The shaded regions are the $1\sigma$ confidence intervals using Monte Carlo sampling of all uncertain parameters. Previous calculations also using the \citet{Shankar2009} BHMF were done in \citet{Kochanek2016} and are plotted in the grey dashed line. Because those calculations were calibrated to an earlier TDE rate measurement from \cite{vanVelzen2014} we re-calibrate them to the \citetalias{Yao2023} local rate, and compare it to our BHMF-only rate (the dashed red line).}
    \label{fig:rate_volumetric}
\end{figure*}

We apply Equation \ref{eq:tde_rate} to calculate a total TDE rate as a function of redshift. In Figure \ref{fig:enhancements}, we show how each scaling independently evolves with redshift. For the density scaling, the shaded region represents the $1\sigma$ confidence interval given uncertainty in the power law scaling of TDE rate with density. For the two BHMFs there is no intrinsic uncertainty so the shaded region shows the same confidence interval propagated from the uncertainty in the local ``baseline'' TDE rate, as measured in \citetalias{Yao2023}. The shaded regions in the merger and obscuration scalings both represent the full parameter range over a uniform  distribution. For the obscuration scaling we find that the exact choices make very little difference on the total number of TDEs observed, meaning even a $90\%$ obscuration fraction at high redshift makes little overall impact.

The primary driver of the TDE rate, as expected, is the decline of the SMBH mass function with redshift. This means that, to zeroth order, previous calculations of the redshift-evolution of the TDE rate are accurate. However, a secondary effect that may dominate at low redshifts is the redshift-evolving density of galaxies. Figure \ref{fig:enhancements} shows that before $z\sim 3$, the galaxy-density enhancement is dominant over the \texttt{Illustris} BHMF downsizing and is comparable to the effect of the \cite{Shankar2009} mass function.  Over $0 \lesssim z \lesssim 2$, the BHMF decrease contributes up to $1.5$~dex of reduction from the local rate, the density enhancement contributes an increase up to $0.5$~dex. Taken together with the galaxy merger rate which contributes an increase up to $0.4$~dex, the total volumetric TDE rate increases from $z=0$ to $z=1$. It turns over at higher redshifts, but this turnover point depends on the BHMF: it is around $z\sim1.2$ for the \citet{Shankar2009} BHMF and $z\sim2$ for \texttt{Illustris}.

The effects of a redshift-evolving IMF and dust obscuration are smaller and insignificant until $z>4$, where the dust obscuration fraction can decrease the high-redshift rate of TDEs up to $0.5$~dex and a top-heavy IMF would increase the high-redshift TDE rate $0.2$~dex. At the highest redshifts we model, $z\sim6$, the typical nuclear density of a galaxy enhances the rate of TDEs by an entire order of magnitude, but the lack of SMBHs capable of disrupting stars ultimately dominates, decreasing the TDE rate to $\sim2 \times 10^{-8}$~Mpc$^{-3}$~yr$^{-1}$ at $z=6$.

We plot the resulting volumetric TDE rate as a function of redshift in Figure \ref{fig:rate_volumetric} for both choices of BHMF, and compared to the redshift-dependent rate from \cite{Kochanek2016}. \cite{Kochanek2016} uses the BHMF from \cite{Shankar2009}, but a different local baseline rate \citep[an earlier study using Pan-STARRS, ][]{vanVelzen2014}. They also assume a black hole mass dependence of the rate, and a luminosity which is dependent on black hole mass (Eddington limited, which is roughly what is observed in optical TDE flares \citep{Mummery2025}), such that their rate differs slightly in its calculation from this work. Their rate largely agrees with this work's BHMF-only calculation, seen in the agreement between their redshift-dependent rate and our \citet{Shankar2009}-based predictions. However, we show that considering another BHMF model completely changes this volumetric rate. Furthermore, given the higher central compactness of galaxies and higher merger rate at cosmic noon, we expect an overall increase in the volumetric rate of TDEs up to $z=2$ before the BHMF definitively takes over and decreases the rate. This effectively extends the results of \citet{Kochanek2016} to include galaxy evolution-dependent effects, as is suggested by the environmental effects discussed in \citet{Pfister2019}, \citet{Pfister2020} and \citet{Hannah2025a}.

Upcoming wide-field and deep surveys will make the first discoveries of populations of (non-jetted) TDEs beyond $z\sim1$, and potentially even very high-redshift $z>3$ TDEs. These populations will be direct probes of the redshift evolution of the TDE rate, and hence the low-mass end of the SMBH mass function. We focus here on the Legacy Survey of Space and Time \citep[LSST,][]{Ivezic2019} conducted using the Vera C. Rubin Observatory (hereafter Rubin), the Roman Space Telescope (hereafter Roman) High Latitude Time Domain Survey \citep[HLTDS,][]{Rose2021, ROTAC2025}, High Latitude Wide area Survey \citep[HLWAS,][]{ROTAC2025}, and the COSMOS-Web survey \citep{Casey2023} conducted with JWST. In this section, we predict TDE detection rates using flux-limited searches in each survey, and propose tests of the BHMF using the observed TDE populations.

\subsection{Legacy Survey of Space and Time (LSST)}

\begin{figure*}
    \centering
    \includegraphics[width=\textwidth]{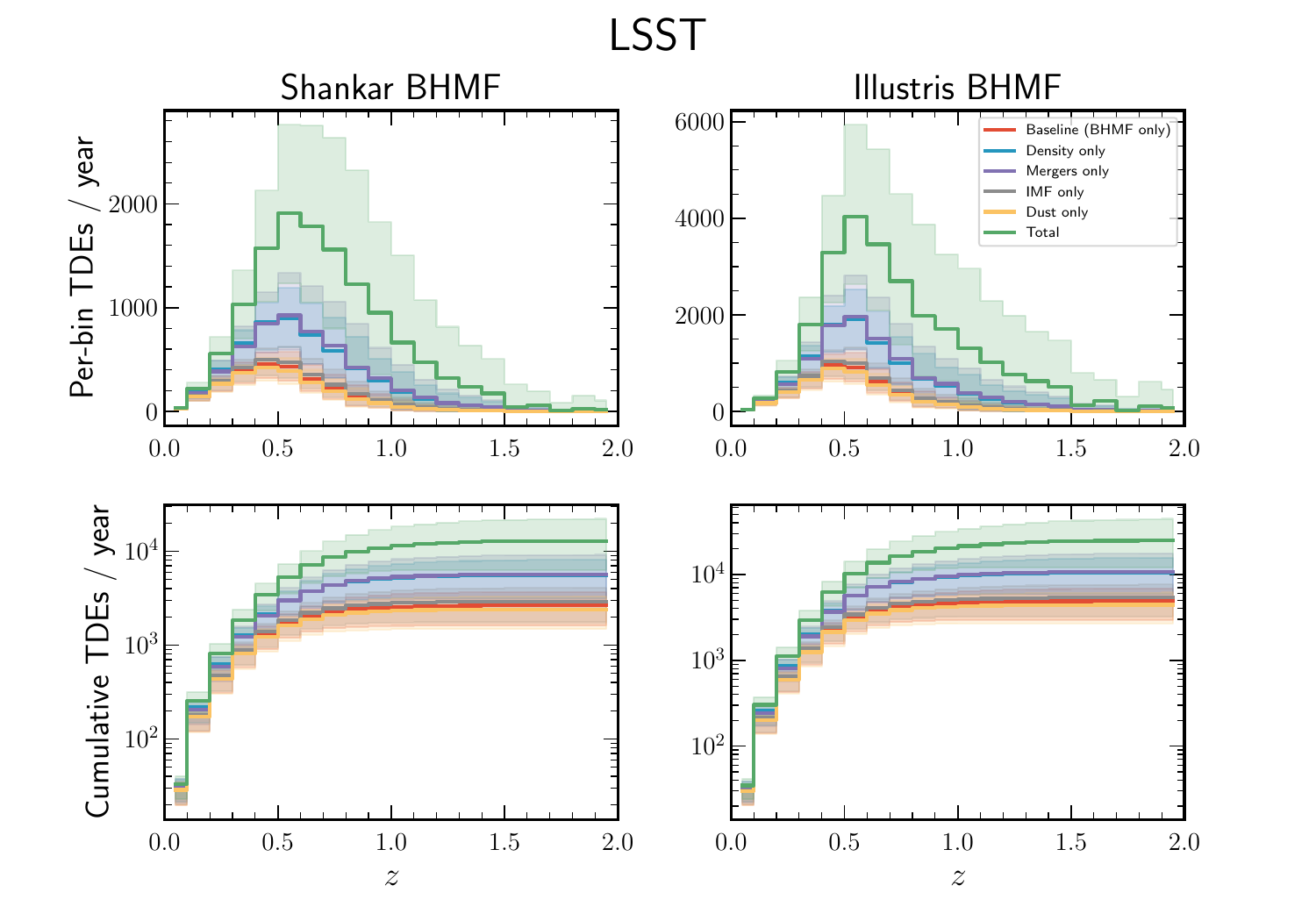}
    \caption{Predicted \textit{observed} rates of TDEs in the Vera Rubin Observatory LSST. The left two panels show rates assuming a \citet{Shankar2009} BHMF, and the right two use the \texttt{Illustris} simulation BHMF. The top two panels show TDEs per redshift bin, and the bottom two show the total cumulative TDEs up to redshift $z$. Each line shows a different single redshift-dependent rate modification (including the BHMF), and the green line shows them all combined.}
    \label{fig:rubin}
\end{figure*}

The LSST survey, conducted by the Vera C. Rubin Observatory, will detect the largest number of astronomical transients ever observed. It is a wide ($\sim 18,000~$deg$^2$) deep ($\sim 25$ AB mag in a single g-band detection) survey, in optical $ugrizy$ filters. The cadence is still being refined at the time of this work, and is ultimately dependent on stochastic factors, but will correspond roughly to one observation every $\sim5$~days \citep{Bianco2022}. 

To account for the potentially poor sampling of the light curves of the faintest, highest-redshift TDEs, we consider a flux-limited survey conducted using LSST, $>1$~mag brighter than the nominal $30$~s single-visit depth in a given LSST filter. This is similar to the flux-limited spectroscopic sample of TDEs from ZTF \citepalias{Yao2023}, and reduces sensitivity to cadence details for very faint sources. We assume that Rubin observes all year long, and do not account for gaps due to weather, maintenance, etc. This allows us to negate those portions of the ``loss factor'' of the survey \citep[see, e.g.][]{Perley2020}, which will decrease our estimates. We find that, at the characteristic redshifts ($\langle z \rangle \approx 0.77$) probed by the TDEs discovered by this flux-limited survey, the $\sim2,000$~\AA{} peak of a $\sim15,000$~K TDE is redshifted to $\sim 3500$~\AA{}. Therefore we report rates in the g band, which yields the largest number of TDE detections of any filter.

In this flux-limited sample, LSST is expected to detect thousands to tens of thousands of TDEs per year. Due to LSST's optical wavelength coverage, very few TDEs are expected to be detected beyond $z\gtrsim1.5$. While this redshift range is much smaller than that of Roman or JWST, the total number of TDEs detected is very sensitive to the low-mass end of the SMBH mass function. As shown in Figure \ref{fig:rubin} and Table \ref{tab:tde_rates}, different BHMF models predict a factor of $\sim2$ difference in the annual rate of TDEs. In Figure \ref{fig:rubin}, the non-monotonic bumps in the rates at $z>1.5$ are a product of the Monte Carlo sampling of the temperature distribution, rather than a physical effect. Additionally, we include predictions for rates in the unstacked deep-drilling fields in Table \ref{tab:tde_rates}, also given a flux-limited survey $1$~mag above the detection limit. We note that the deep-drilling fields will be able to probe even deeper than listed given coadded observations.

Previous calculations of the TDE yield of LSST \citep{Brijcman2020} do not evolve the black hole mass function with redshift and therefore systematically predict higher rates ($3500-8000$~TDEs/year) than the BHMF-only prediction in this work ($2500-5000$~TDEs/year, depending on the BHMF). It should be noted that \citet{Brijcman2020} require $\geq 10$ detections, while in this work we require only a single detection $\geq1$~mag above the LSST limit in any filter.

%Our results show that LSST provides a powerful constraint on the normalization of the evolving TDE rate, and therefore the integrated number density of lower-mass SMBHs.

It is important to distinguish between the number of TDEs detected (a single detection in any filter) and the number of TDEs robustly classified and discovered in LSST. \cite{Bricman2023} performed a study injecting model TDE light curves into a proposed LSST cadence. They find that only $7.5\%$ of TDEs with any detections fulfill their discovery metric, in which measurements of color evolution enable TDEs to be distinguished from supernovae (SNe). They also find that higher-redshift samples of TDEs are biased against faster-evolving events, such as iPTF16fnl \citep{Blagorodnova2017}. The cadence they use is an earlier simulation of the LSST cadence, v2.1 rather than current (at the time of writing this paper) v5.0.0. Changes to the cadence include changes to exposure times, readout times, footprint, and general strategy all of which could make a significant impact on the figure of merit calculated in \cite{Bricman2023}. Nevertheless, their results provide a reasonable estimate of LSST’s classification efficiency. 

Because our analysis adopts a shallower, flux-limited selection than the
full LSST alert stream, we expect a higher completeness for rapidly evolving
TDEs than implied by these studies. Given these selection effects, perhaps ``only'' thousands of TDEs will be classified photometrically each year-- and tens at $z>1$ given blind searches. 

LSST's strength is in constraining the overall normalization of the TDE rate at low and intermediate redshifts. This anchors the rates calculated from Roman and JWST TDE samples, allowing those surveys to more effectively discriminate between SMBH mass function models. In Section~\ref{sec:testing}, we show that even coarse statistical summaries of the LSST TDE population — such as the total yield and median redshift — can already place meaningful constraints on the redshift evolution of the low-mass SMBH mass function.

\subsection{Roman Space Telescope}

The High Latitude Time Domain Survey (HLTDS), conducted with the Roman Space Telescope, is split into two tiers: a wide ($\sim 19~\mathrm{deg}^2$) tier and a deep ($\sim 6~\mathrm{deg}^2$) tier \citep{Rose2021}. In this work we primarily focus on the wide tier, which is expected to discover order(s) of magnitude more TDEs and will therefore provide the strongest statistical constraints on the redshift evolution of the TDE rate. 

The Roman Observations Time Allocation Committee (ROTAC) report \citep{ROTAC2025} provides the exposure times, cadences, and filters for the HLTDS. We use these, along with the example notebook in the Roman Exposure Time Calculator (ETC) to calculate $5\sigma$ point source depths, to the nearest $0.05$~AB mag (always rounded down, thus slightly shallower than the actual survey). We assume the HLTDS magnitude limits to be $25.95$~mag in the $F062$ filter, 25.05~mag in $F087$, 25.20~mag in $F106$, $25.65$~mag in $F129$, and $26.10$~mag in $F158$. Among these filters, $F062$ will yield the highest TDE rate due to its depth and alignment with the observed wavelength of higher-redshift TDEs, and we therefore report rates for this filter.

% We also perform calculations for TDEs serendipitously detected in a single epoch of the High Latitude Wide Area Survey (HLWAS) in Table \ref{tab:tde_rates}.

\begin{figure*}
    \centering
    \includegraphics[width=\textwidth]{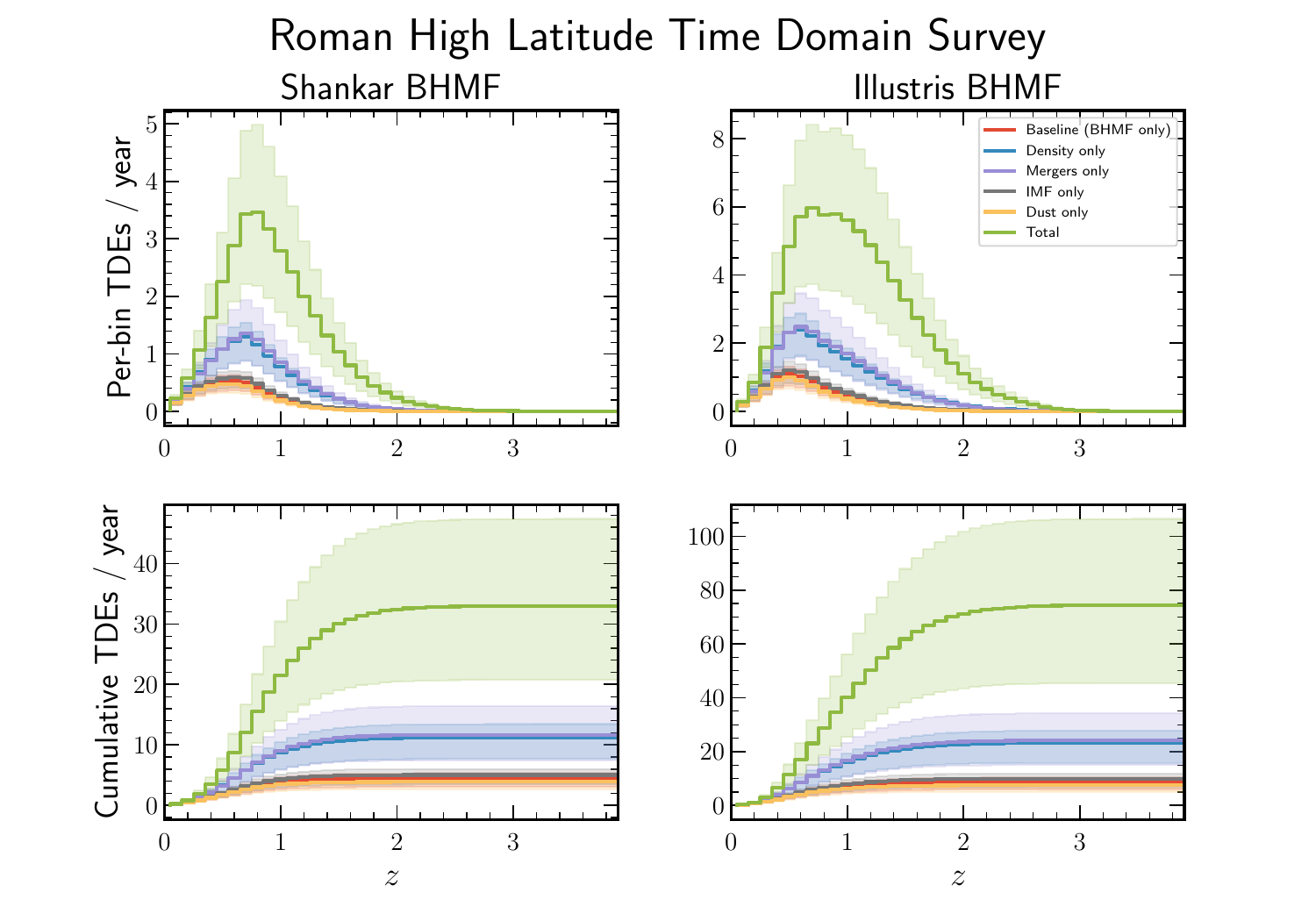}
    \caption{Predicted \textit{observed} rates of TDEs in the Roman High Latitude Time Domain Survey, as described in Figure \ref{fig:rubin}.}
    \label{fig:roman}
\end{figure*}

Figure \ref{fig:roman} shows the predicted observed TDE rates in the HLTDS wide tier. We report the rates for both tiers of the HLTDS in Table \ref{tab:tde_rates}. In the wide tier, we find that the annual TDE yield will range from several dozen to $\sim 100$ TDEs per year, depending on the redshift-dependent enhancements. While the absolute number is smaller than in LSST, the Roman TDE sample is expected to be exceptionally clean and well-characterized. The cadence and filter set of the HLTDS is ideal for a very complete sample of $z>1$ TDEs: at these higher redshifts, optical TDE emission is k-corrected into the redder and NIR filters that the survey utilizes. It is equipped to discover TDEs up to $z\sim3$ if high-redshift luminous TDEs occur within the field of regard during the survey, however they will be quite rare. The HLTDS will sensitively probe both the normalization of the TDE redshift distribution and its shape. Therefore, the HLTDS will be a powerful discriminator between BHMF models, especially when forward-modeled redshift distributions are compared using statistical tests such as Kolmogorov–Smirnov or likelihood analyses (see Section~\ref{sec:testing}). 

We also include the predictions for the number of TDEs observed serendipitously by a single epoch of the Roman High Latitude Wide Area Survey (HLWAS), although it is uncertain how such TDEs could be discovered by limited time-series data. In a single exposure, it is possible to observe TDEs out to $z>4$, and thousands of TDEs will be detected. However, the highest-redshift TDEs will appear as faint, hostless point sources \citep[see discussion in][]{Karmen2025} and will be extremely difficult to classify.

The high spatial resolution imaging that Roman provides will allow a robust understanding of the TDE host morphologies, e.g. compactness, central density, and merger signatures. This can be used to disentangle TDE rate dependencies on the SMBH mass function from the effects of host galaxy properties. Recent modeling frameworks such as REPTiDE \citep{Hannah2025b} are already able to relate the galaxy profile to the underlying TDE rate. With a Roman-quality sample, this can be applied systematically allowing a joint analysis of both the underlying black holes, and the TDE hosts. 

Alongside discoveries of higher-redshift TDEs, Roman will also supplement the optical coverage of lower-redshift TDEs with near-infrared observations. This will extend the coverage of the spectral energy distribution (SED) of many TDEs into the near-IR, providing further information about their emission. In particular, the LSST deep drilling fields will overlap with the Roman High Latitude Time Domain Survey, providing near-IR monitoring of a large number of TDEs optically discovered by LSST. Additionally, Roman has the capability of discovering dust echoes from TDEs, particularly the hottest components. Infrared echo studies of optically discovered TDEs show that UV/optical emission is reprocessed by circumnuclear dust on sub-parsec scales, producing mid-infrared variability on timescales of months to years that can be robustly identified through difference imaging \citep{Jiang2021}. These echoes typically have peak luminosities of $\sim10^{41}$--$10^{42}\,{\rm erg\,s^{-1}}$ and arise from dust located at $\lesssim0.2$ pc, with low covering factors ($f_c\sim0.01$ or less), implying that many optically selected TDE hosts contain relatively little nuclear dust. Roman’s near-infrared sensitivity will primarily probe the hotter, earlier phases of this reprocessed emission, enabling detection of nuclear IR variability contemporaneous with or shortly following the optical flare.

\subsection{JWST COSMOS-Web}

\begin{figure*}
    \centering
    \includegraphics[width=\textwidth]{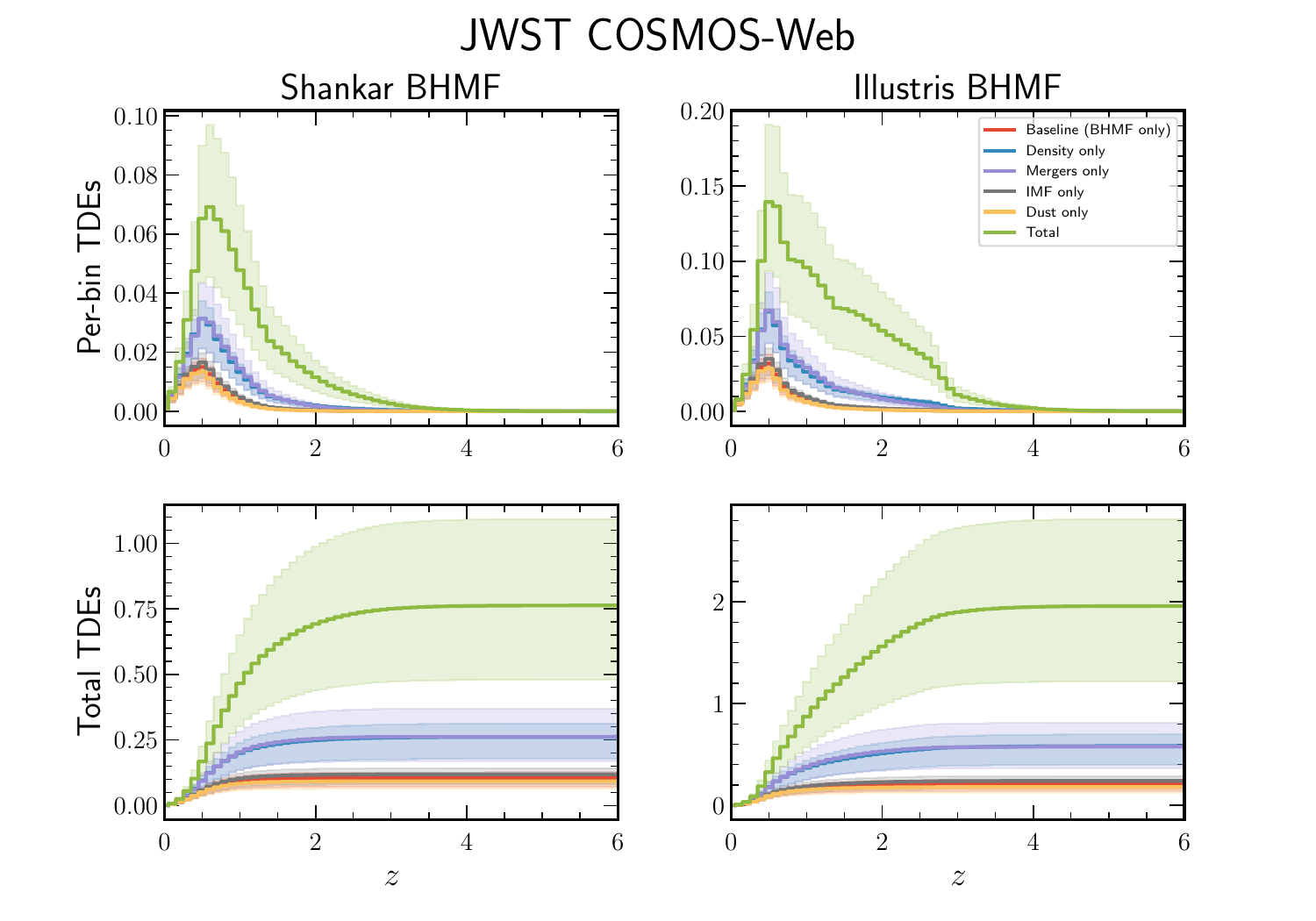}
    \caption{Predicted \textit{observed} rates of TDEs in the JWST COSMOS-Web survey. Rates are as described in Figure \ref{fig:rubin}.}
    \label{fig:cosmos}
\end{figure*}

JWST could detect a typical TDE out to $z\sim10$ \citep{Karmen2025}. A range of theoretical models predict dramatic changes in galaxy structure, stellar populations, and black hole growth at these early cosmic times, all of which would have dramatic implications for the TDE rate \citep{Pfister2021,  Vanzella2023, Baggen2023, Inayoshi2024, Guia2024, Adamo2024, Chowdhury2024, Kritos2025}. Therefore, a single confirmed high-redshift TDE would place powerful constraints on SMBH seeding, nuclear stellar populations, and compact stellar systems. 

Among the deep, wide-field surveys conducted with JWST, COSMOS-Web has the widest area ($0.54~\rm{deg}^2$) which gives it the highest probability of detecting a TDE serendipitously. COSMOS-Web observes in four NIRCam filters, with $5\sigma$ depths of 26.87 mag in $F115W$, 27.14 mag in $F150W$, 27.71 mag in $F277W$, and 27.61 in $F444W$ \citep{Casey2023}. We predict that  given these depths, depending on the BHMF, COSMOS-Web may serendipitously capture between zero and two TDEs in a single epoch as seen in Figure \ref{fig:cosmos}. Most TDEs detected through this method would lie at $z<2$, with very small chances of any TDEs beyond $z>2$ regardless of the BHMF model.

Given the filters and limiting magnitudes of COSMOS-Web, the distance to which a TDE can be detected is limited ultimately by the Lyman alpha forest, which absorbs the majority of light in the $F115W$ filter beginning at $z\sim7.2$. At these extreme redshifts, the $\sim1,500$~\AA{} blackbody peak of a $20,000$~K TDE's emission is k-corrected into the $F115W$ filter, meaning that at these distances NIRCam is observing the same rest-frame optical emission that is traditionally used to discover TDEs from ground-based observatories. Similarly, time dilation stretches the $\sim2$-week rise to $\sim4$~months, and the $\sim2$~month fade to $2$-years. This makes a deep galaxy survey such as COSMOS-Web, which has its observations spread out over $\sim$a year and nearly two decades of historic imaging, somewhat well-suited toward the discovery of rest-UV-bright transients.

The TDE rate in the COSMOS field is limited, ultimately, by its tiny area and (lack of) cadence. While JWST is not designed to conduct wide field surveys (such as the Roman High Latitude Wide Area Survey), repeated observations of deep extragalactic fields can effectively provide both ultra-deep co-added imaging for galaxy evolution science and difference imaging suitable for transient discovery. Yearly observations of a given field, rather than a single epoch, may be the key to discovering a $z>3$~TDE in JWST. For example, we calculate that given two extra epochs of imaging in the COSMOS field, we expect $N\approx1$ TDEs at $z\geq3$. Already, a search of the $0.037~\rm{deg}^2$ overlap between the PRIMER and COSMOS-Web observations yielded the discovery of 68 supernovae \citep{Fox2026}. While within such a small area one only expects $N\approx0.07$~TDEs. Thus, long-term variability monitoring strategies, such as those which have been suggested for JWST \citep{STScI_LTVM} would be ideal for finding the highest-redshift TDEs. 

JWST therefore serves a different role than LSST or
Roman. Rather than providing statistical constraints on the TDE rate,
JWST offers access to the extreme tail of the redshift distribution, where individual TDE detections can directly test models of early black hole growth and dense stellar systems. Detailed strategies for identifying high-redshift TDEs in JWST data are discussed in \cite{Karmen2025}.

\begin{deluxetable*}{llcccc}
\tablecaption{Forecasted TDE yields and redshift distributions. For all predictions we use the TDE rate which accounts for all galaxy morphology effects, and the BHMF. For LSST and the Roman HLTDS, we calculate the one-year yields and in COSMOS-Web and the Roman HLWAS we calculate the number of TDEs serendipitously observed by a single epoch.}
\label{tab:tde_rates}
\tablehead{
Survey & BHMF & $N_{\rm TDE}$ & $z_{\rm med}$ & $\langle z\rangle$ & $z_{\max}(N_{\rm TDE}\ge1)$
}
\startdata
JWST COSMOS-Web & Illustris/TNG & $1.91^{+0.89}_{-0.71}$ & $1.17^{+0.04}_{-0.08}$ & $1.40^{+0.03}_{-0.05}$ & -- \\
JWST COSMOS-Web & Shankar+09 & $0.76^{+0.33}_{-0.27}$ & $0.89^{+0.02}_{-0.05}$ & $1.12^{+0.03}_{-0.04}$ & -- \\
Rubin (LSST) & Illustris/TNG & $27094^{+11713}_{-9985}$ & $0.64^{+0.06}_{-0.04}$ & $0.77^{+0.04}_{-0.05}$ & $1.95^{+0.00}_{-0.20}$ \\
Rubin (LSST) & Shankar+09 & $13849^{+5956}_{-4877}$ & $0.63^{+0.05}_{-0.04}$ & $0.73^{+0.04}_{-0.04}$ & $1.95^{+0.00}_{-0.20}$ \\
Rubin (LSST deep drilling) & Illustris/TNG & $98.76^{+42.82}_{-35.04}$ & $0.70^{+0.05}_{-0.05}$ & $0.87^{+0.04}_{-0.05}$ & $2.35^{+0.20}_{-0.20}$ \\
Rubin (LSST deep drilling) & Shankar+09 & $48.76^{+19.85}_{-16.86}$ & $0.67^{+0.04}_{-0.04}$ & $0.78^{+0.03}_{-0.04}$ & $1.85^{+0.20}_{-0.30}$ \\
Roman HLTDS (deep tier) & Illustris/TNG & $31.12^{+14.21}_{-11.81}$ & $1.22^{+0.04}_{-0.05}$ & $1.36^{+0.04}_{-0.05}$ & $2.75^{+0.10}_{-0.30}$ \\
Roman HLTDS (deep tier) & Shankar+09 & $12.23^{+5.46}_{-4.61}$ & $0.99^{+0.02}_{-0.04}$ & $1.14^{+0.03}_{-0.04}$ & $1.95^{+0.20}_{-0.20}$ \\
Roman HLTDS (wide tier) & Illustris/TNG & $73.93^{+32.99}_{-27.52}$ & $1.00^{+0.07}_{-0.07}$ & $1.14^{+0.05}_{-0.06}$ & $2.65^{+0.10}_{-0.30}$ \\
Roman HLTDS (wide tier) & Shankar+09 & $32.39^{+14.65}_{-11.66}$ & $0.87^{+0.04}_{-0.04}$ & $0.99^{+0.04}_{-0.04}$ & $2.05^{+0.20}_{-0.20}$ \\
Roman HLWAS (single-epoch) & Illustris/TNG & $4916^{+2159}_{-1822}$ & $1.05^{+0.06}_{-0.10}$ & $1.26^{+0.04}_{-0.06}$ & $4.50^{+0.00}_{-0.20}$ \\
Roman HLWAS (single-epoch) & Shankar+09 & $1995^{+832}_{-725}$ & $0.78^{+0.03}_{-0.07}$ & $1.02^{+0.03}_{-0.06}$ & $4.30^{+0.20}_{-0.40}$ \\
\enddata
\end{deluxetable*}

\subsection{Zwicky Transient Facility}

\begin{figure*}
    \centering
    \includegraphics[width=\linewidth]{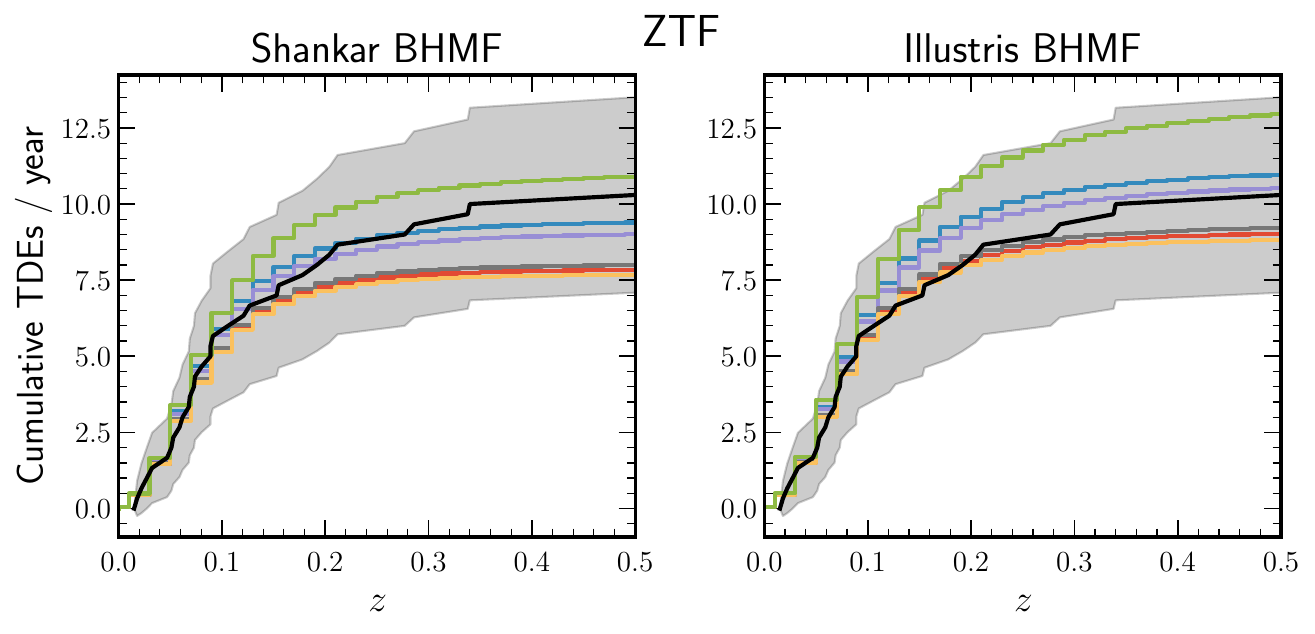}
    \caption{Cumulative redshift distribution of TDEs as observed by ZTF. The black line is the observed TDEs from the $\sim3$~year ZTF flux-limited sample in \citetalias{Yao2023}, and the surrounding shaded region is the Poisson uncertainty. The filled curves are our TDE rates models, scaled using the loss factor of ZTF. While the models differ slightly, they all fall within the ZTF uncertainty region rendering them indistinguishable.}
    \label{fig:ztf}
\end{figure*}

Given the different predictions of redshift-dependent rate modifications, one may ask if these differences are already seen in ZTF. We show in Figure \ref{fig:ztf} the observed cumulative redshift distribution in ZTF, along with our modeled TDE rates. As expected, these align almost exactly because we calibrate all of our rate predictions to the observations from ZTF. However, it is reassuring that the shape of the cumulative distribution function (CDF) matches the observed shape, which is not calibrated into our model. This confirms that the only factor influencing the redshift distribution of the ZTF-observed TDEs is the volume probed by the survey.  We find that none of the models are distinguishable using the current ZTF sample. The two BHMF models only differ in their TDE rate at $\sim 5\%$, so it would take a sample size of $\sim400$ TDEs (assuming the same magnitude limit of $19.5$~mag, and Poisson statistics) to observe this divergence. However, the models become more easily distinguishable in deeper surveys that probe higher-redshift TDEs, such that the models more strongly differ.

\subsection{Tests of the BHMF with LSST}
\label{sec:testing}

\begin{figure}
    \centering
    \includegraphics[width=\linewidth]{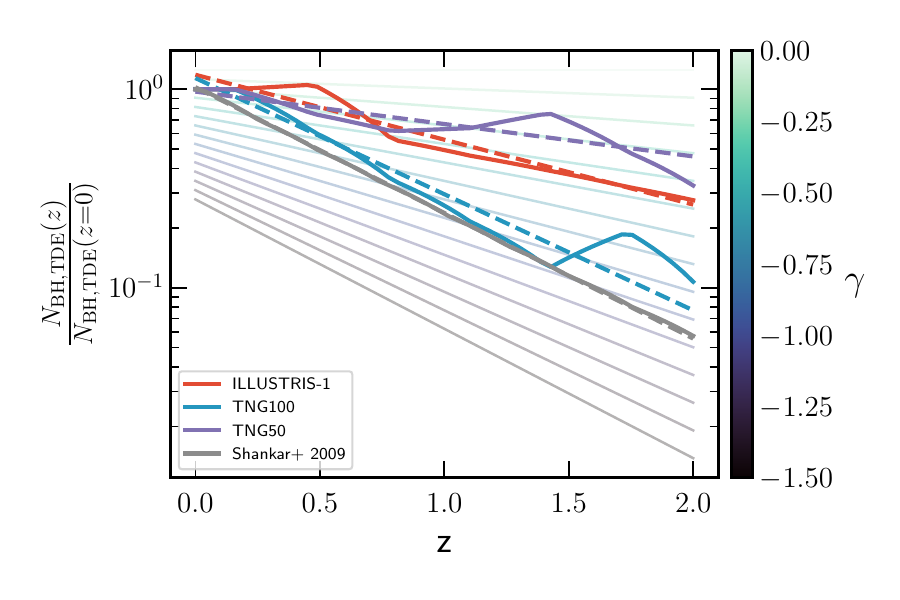}
    \caption{Exponential approximations for the black hole mass function ``slope'', colored by exponent $\gamma$ (Equation \ref{eq:downsizing}). The semiempirical BHMF \citep{Shankar2009} is well-approximated by the exponential downsizing with redshift, while the simulated BHMFs behave more stochastically. }
    \label{fig:exponential_bhmf}
\end{figure}

Ideally, the absolute number of TDEs in a flux-limited survey (if it samples TDEs out to $z\gtrsim1$) can be a robust test of the low-mass end of the black hole mass function's evolution with redshift (hereafter, when we refer to the BHMF we are referring to the redshift evolution of the low-mass end). It can be seen in Equation \ref{eq:tde_rate} and Figures \ref{fig:enhancements} and \ref{fig:rate_volumetric} that the BHMF is a primary driver of the TDE rate's dependence on redshift. The largest difference between the predictions from the two BHMF models we use is the number of TDEs detected per year in a given survey. For the Roman Space Telescope's HLTDS, the detection efficiency will be very high and galaxy morphology can be well-measured so this number will be constraining. However, for LSST, the detection efficiency is yet unknown and highly dependent on overheads, weather, moon phase, etc. Ideally the survey efficiency will be well-measured and corrected for, thus we can put a first constraint on the BHMF by using this quantity.

With the number of TDEs detected in a given year, we can test any given BHMF using LSST. To illustrate this, we approximate the ``slope'' (i.e., the evolution with redshift) of an arbitrary BHMF within the TDE BH mass range, $N_{\rm BH}(z)$ (as defined in Equation \ref{eq:n}) as an exponential decline, fit as 
\begin{equation}
    N_{\rm BH}(z) \approx Ae^{\gamma (1+z)}
    \label{eq:downsizing}
\end{equation}

For the \texttt{Illustris} BHMF (which declines more slowly in the low-mass end) we find $\gamma = -0.82$, and for the Shankar BHMF (which declines more rapidly) we find $\gamma = -1.46$. $\gamma=0$ corresponds to a static TDE rate throughout the history of the Universe. These approximations are shown in Figure \ref{fig:exponential_bhmf}, alongside fits to the TNG100 and TNG50 simulations \citep{Habouzit2021}. Using this exponential approximation rather than the actual BHMF model reproduces the predicted total number of TDEs to within $1\%$. 

\begin{figure*}
    \centering
    \includegraphics[width=\linewidth]{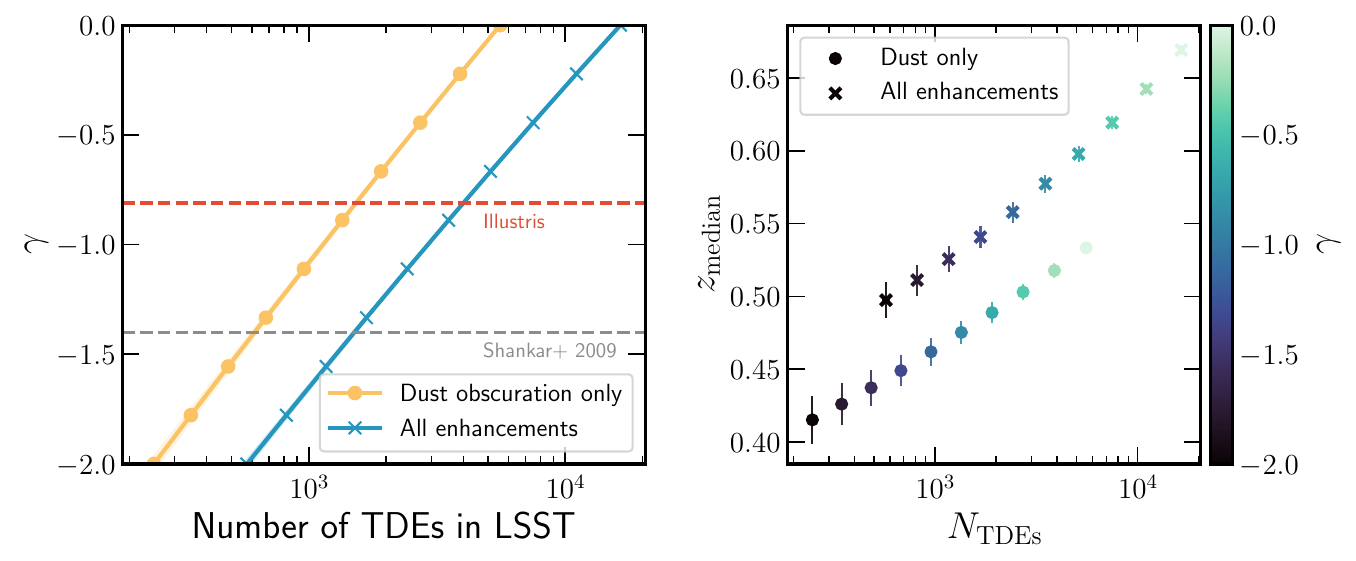}
    \caption{\textbf{Left:} The number of TDEs detected in a flux-limited survey conducted with LSST vs the exponent index of the mass function evolution (Equation \ref{eq:downsizing}, see Figure \ref{fig:exponential_bhmf}). The lowest prediction is the dust-only calculation (yellow circles), and the highest is the total calculation (blue x's); all predictions of the TDE yield lie between these lines. The exponent indices fit to the \citet{Shankar2009} and the \texttt{Illustris} mass functions are horizontal dotted lines. \textbf{Right}: The number of TDEs detected in a flux-limited survey conducted with LSST vs the median redshift for each sample. The exponent index (y-axis of the left panel) is the color map. Uncertainties in sample medians are calculated via bootstrap resampling.}
    \label{fig:tests}
\end{figure*}

We plot the exponent index $\gamma$ against the total number of TDEs detected in the left-hand panel of Figure \ref{fig:tests}. As expected, we see that the relationship is logarithmic (linear in $\gamma$ vs $\log(N_{\rm TDE})$), and the number of detected TDEs decreases as the BHMF slope becomes steeper ($\gamma$ decreases). The uncertainties increase for more rapidly evolving BHMFs because of the smaller number of total TDEs detected. We incorporate the other redshift-dependent TDE rate enhancements to see if the evolution of the BHMF is still measurable.  This is also seen in Figure \ref{fig:tests}; between the lowest prediction of the TDE rate, in which only the dust obscuration fraction changes with redshift, and the highest prediction in which all redshift-dependent effects occur, we have a range of $0.3$~dex. This means, using only the total number of TDEs detected, one can constrain $\gamma$ to within $\pm 0.6$ due to the degeneracy between galaxy-scale effects and BHMF effects. 

One can further discern the effects of the BHMF from galaxy-scale effects from the redshift distribution of the observed TDEs. In the right-hand panel of Figure \ref{fig:tests} we show the number of TDEs detected vs the median redshift of a given TDE sample. Here, the galaxy-scale enhancements and BHMF result in different sample median redshifts for the same annual TDE yield. We calculate an uncertainty in the sample median by bootstrap resampling a smoothed version of the LSST redshift distribution, and assuming the yield of a $1$~year survey. While the difference in median redshift for a given TDE yield is only $0.05$, it is significant given the small statistical errors. Now, given the annual TDE yield and median sample redshift (and assuming local linearity, $\gamma \approx a + b\log(N_{\rm TDE}) + c z_{\rm median}$) we get a mean uncertainty in BHMF slope $\gamma$ of $0.06$. Overall, a well-understood flux-limited LSST survey, combined with basic statistical measurements of the TDE redshift distribution, can place strong constraints on the redshift evolution of the BHMF for $M_{\bullet} \leq 10^8 M_{\odot}$. 

For both LSST and Roman, the strongest but most expensive measurement of the BHMF is a full forward model of the TDE redshift distribution using methods such as those presented in this work. The forward model can then be compared to the observed redshift distribution using a Kolmogorov–Smirnov (K-S) test. For the Roman HLTDS TDE sample, which is extremely sensitive to the BHMF and galaxy-level effects, the K-S test is the ideal probe of the BHMF. Within an individual redshift bin, the peak luminosity of the TDE optical flare also correlates with the SMBH mass \citep{Mummery2024}, and can further constrain the distribution of black hole masses disrupting stars. This is further explored in \cite{Ramsden2025} and can be used to understand the mass function at a given redshift.

\section{Discussion}
\label{sec:discussion}

\subsection{The redshift-evolving rate of TDEs}

TDE rates reflect the balance between the supply of stars in the nuclei of galaxies, and the demographics of central supermassive black holes. On one side, there must be enough galaxies hosting SMBHs in the mass range capable of visibly disrupting stars. On the other, a dense stellar population is needed near the central SMBH for disruptions to occur at an appreciable rate. 

In the local universe, this interplay is most clearly seen in PSB and compact green-valley galaxies, which are strongly over-represented among optically selected TDE hosts \citep{LawSmith2017, French2020, Hammerstein2021}. These systems have low enough masses to host SMBHs which are below the Hills mass, but have undergone recent star formation or structural reconfiguration that leave behind dense nuclear stellar populations. 

At cosmic noon, these rare green-valley systems-- particularly PSB galaxies-- are much more prevalent \citep{Wild2016} and SMBHs have undergone most of their growth since their high-redshift seeds \citep{Inayoshi2020}. As a result, it is not unexpected that the volumetric TDE rate increases toward $z\sim2$, until the declining number density of SMBHs that can disrupt stars ultimately dominates and drives the rate lower at higher redshifts. 

 The location of this turnover, as seen in Figure~\ref{fig:rate_volumetric}, depends sensitively on the BHMF as well as the strength of galaxy-scale enhancements. However, when controlling for host galaxies, as will be the case in Roman (or LSST after many years of coadded imaging, this redshift-dependent rate becomes a sensitive probe of SMBH growth that is complementary to studies of high-redshift AGN. It should be noted that up to 50\% of TDEs may be obscured by dust at $z=2$, which makes mid-infrared searches for dust emission such as that done in \citet{Masterson2024} essential to complete the optically-selected sample of TDEs.

\subsection{Limitations of the Method}
\label{sec:limitations}
In this work, we assume that the shape of the luminosity function of TDEs is constant with redshift (only its normalization changes, given our derived scalings), but the black hole mass function evolves with redshift. These two assumptions are fundamentally incompatible, because the peak optical/UV luminosity of a TDE is seen to depend on its SMBH mass \citep{Mummery2024, Yao2023}. However, given the observed scatter in this dependence, we do not expect the luminosity function to be extremely sensitive to the mass function. Even a major change in the mass function should not effect a major change in the median TDE luminosity at higher redshifts. With redshift, all BHMF models become more bottom-heavy. Therefore it is only the most massive (and therefore, most luminous) end of the BHMF and thus luminosity function that is affected by this evolution. Because this end holds the least density in the luminosity function, it only causes an effect of a few percent, as seen in the difference between our BHMF-only volumetric rate and the \cite{Kochanek2016} rate at $z\geq 4$ (see Figure \ref{fig:rate_volumetric}). Taking this limitation into account, we only make predictions about the number of TDEs detected and not their apparent magnitude distribution or black hole mass distribution.

However, the most rigorous way to do this prediction would be to parameterize everything in terms of BH mass. One can add the observed random scatter when modeling the mass-dependent luminosity, and then follow our methodology given a redshift-dependent luminosity function. In this work, we do not perform this analysis due to the large uncertainty in the mass-dependent TDE rate. The local SMBH mass function is measured, and the local TDE BH mass function is observed, so these two functions relate via the TDE rate as a function of BH mass. This is proposed to be anywhere from $\Gamma \propto M_{\bullet}^{-0.25}$ \citep{Alexander2017} to $\Gamma \propto M_{\bullet}^{-0.4}$ \citep{Stone2016}. In more recent semi-theoretical work, it is proposed that this relation may be even steeper for high-mass BHs, with the inverse power-law relation below $M_\bullet < 10^6~M_\odot$ \citep{Chang2025, Hannah2025b}. To observationally measure this, it must be disentangled from any black-hole-mass-dependent selection effects that affect the observed TDE rate and observed SMBH mass function. These include lack of TDE searches in active AGN, requirements for visible host galaxies, and other effects that affect the mass range of SMBHs which cause TDEs. Addressing these selection effects is beyond the scope of this work.

It has been shown \citep{Mummery2024} that the late-time UV/optical plateau is well-correlated with SMBH mass. Fitting a compact accretion disk to this late time UV emission combined with X-ray observations is shown to find the SMBH mass most consistent with other mass-scaling relations ($M_{\bullet}-\sigma$ and $M_\bullet-M_{\star}$) \citep{Guolo2025}. This late time plateau is therefore incredibly powerful at discerning SMBH mass, with scatter in this model driven by the assumed distribution for remaining free parameters of the model: inclination, spin, initial disk mass (or stellar mass) and radius, penetration factor and effective viscosity, all of which are subdominant \citep[see Figure 6 of][]{Mummery2024} with respect to $M_\bullet$. However, the dependence on late time plateau means that observations are reliant on emission that is $10-100\times$ fainter than the peak emission of the TDE. Given smaller samples such as the ZTF TDE sample, it is reasonable to systematically follow up TDEs with late time deep UV/optical imaging and measure the plateau. However, this follow-up is infeasible given the potentially thousands of TDEs discovered every year by LSST; LSST will only serendipitously observe the brightest of these plateaus. This reduces the total volume of SMBHs probed by TDEs by 2 to 3 orders of magnitude. Therefore, although less precise in individual measurements, analysis of the redshift distribution of the populations of TDEs in upcoming surveys will offer a much larger volume within which we can study the masses of SMBHs. 

We also re-calculate our predictions extending the maximum TDE luminosity, $L_{\rm max}$ in Equation~\ref{eq:baseline_rate} to $10^{45.8}~\rm{erg~s}^{-1}$, the maximum luminosity of extreme nuclear transients \citep[ENTs][]{Henkly2025} which have been proposed to have TDE origin \citep[e.g.][]{Subrayan2023}. Due to their high luminosities, it is expected that they can be detected to even greater distances, however because of the dramatic decline in the ZTF luminosity function used in this work beyond $L_g > 10^{44}~\rm{erg~s}^{-1}$, extending the luminosity function has no effect on $z_{\rm max}$ in Table \ref{tab:tde_rates} and minimal effect on $N_{\rm TDE}$. This highlights the failure of traditional TDE search methodologies, such as \citetalias{Yao2023}, to identify ENTs and the need for separate rate calculations to made if these are to be included in the TDE luminosity function. The evolution of the abundance of these events with redshift becomes increasingly involved if \citep[as claimed by e.g.][]{Graham2026}, these events involve very massive ($M_\bullet \geq10^8~M_{\odot}$) SMBHs and/or massive stars. Similarly, the ZTF luminosity function does not include the few jetted TDEs \citep{Brown2015, Pasham2015, Andreoni2022}, and would need modification to make accurate predictions of their rates.

\subsection{COSMOS-Web TDE Search}

As a validation of our COSMOS-Web rate predictions, we look to an actual search for TDEs in COSMOS-Web. We previously performed a search for any $z\geq 3.5$ TDEs serendipitously detected in the COSMOS-Web survey \citep{Karmen2025}. That work showed that TDEs in this specific redshift range will not have detected host galaxies in COSMOS-Web. We therefore searched for point sources that resemble the ultraviolet early-time flare of a TDE. We identified one source that perfectly fits the SED of a high-redshift, hot thermal source and upon comparison with archival COSMOS field imaging, confirmed it is indeed a transient. It has no obvious host galaxy, and find that it is well-fitted by both a $z>4$ TDE, or a $z\sim3$ superluminous supernova.

We have since attained follow-up observations thanks to Director's Discretionary Time on JWST granted by STScI (DDT: 9356, PI: M. Karmen). These observations will be described in further detail in upcoming work (Karmen et al., in preparation). The observations consist of $7.2$~hours of NIRSpec PRISM spectroscopy, $\sim1$ observer-frame year after the initial detection. 
% The observation uses the S400A1 slit and the NRSRAPID readout pattern, with 8 integrations per exposure and 3 dithers. We used low-resolution, deep spectroscopy such that given even a lower-redshift, rapidly-evolving TDE we still expected to be able to detect the transient to $5\sigma$. 
The COSMOS-3D \citep{Kakiichi2024} survey also performed NIRSpec grism spectroscopy with concurrent photometry of the source soon after our observations, adding another data point to the light curve. From these observations, we conclude that the TDE candidate is in fact a lower-redshift supernova with an exceptionally large host-galaxy separation. We will provide further details on this conclusion in future work by Karmen et al. in preparation.

% The transient faded faster than any TDE models predict, and COSMOS-3D photometric non-detection shows an upper-limit of 27.5~mag in F115W. The PRISM spectrum is consistent with this, with only emission lines weakly present and no continuum detected. This implies it is either a transient at a lower redshift or a more rapidly-evolving transient. In (Karmen et al., in prep.) we provide a detailed analysis of the spectrum, but here we briefly summarize. The emission lines are significantly brighter than the surrounding background, so they must be remaining emission from the transient, and not from its host. We fit the strongest lines and find that the transient was at $z=1.8$, which is the same redshift as the photometric redshift assigned to a nearby galaxy. We therefore assume that galaxy is the host, and that this TDE candidate is a very high separation supernova. 

Given the rates of TDEs in the COSMOS-Web footprint that we calculate in this work, it is reasonable that a $z\geq3.5$ TDE was not found in a search of COSMOS-Web. Even the most optimistic models predict a median redshift of $z=1.2$ for any TDEs serendipitously discovered in a single epoch. However, we emphasize that the potential of JWST for discovering high-redshift TDEs scales rapidly with number of epochs, and repeated observations of deep galaxy fields would inevitably lead to the discovery of the most distant TDEs.

\section{Conclusion}
\label{sec:conclusion}
In this work, we develop a semiempirical, redshift-dependent model for the TDE rate. The model incorporates the evolving supermassive black hole mass function, along with several galaxy-scale processes that are expected to change over cosmic time. We begin the model with the empirically measured local TDE luminosity function from ZTF, and forward-model the effects of redshift evolution in black hole demographics, galaxy structure, mergers, obscuration, and stellar populations. Using this framework, we generate predictions for upcoming time-domain and deep-field surveys.

\begin{itemize}
    \item We show that the volumetric TDE rate at $z \gtrsim 0.5$ is strongly regulated by the evolution of the low-mass ($10^5$–$10^8~M_\odot$) end of the SMBH mass function, which remains poorly constrained by AGN-based measurements. Plausible BHMF models calibrated to existing data, such as the semiempirical model of \citet{Shankar2009} and the \texttt{Illustris} simulations \citep{Genel2014, Sijacki2015}, predict TDE rates that differ by up to an order of magnitude by $z\sim3$. Thus, TDE rates will provide a probe of SMBH growth in a mass and redshift range otherwise inaccessible.
    \item Galaxy-scale processes, namely increasing nuclear stellar densities, enhanced galaxy merger rates near cosmic noon, and dust obscuration, can have powerful effects on the intermediate-redshift TDE rate. At $z\lesssim 2$, these effects may overcome the BHMF and result in a volumetric rate of TDEs which increases until cosmic noon.
    \item LSST will detect thousands to tens of thousands of TDEs per year in a flux-limited sample. This sample can provide strong constraints on the redshift evolution of the TDE rate, and thus the underlying BHMF. 
    \item The Roman HLTDS will detect fewer TDEs, but with higher completeness, extended redshift reach ($z=2.75$), and high-resolution host-galaxy imaging. This leads to a complementary test of both black hole demographics and host galaxy-driven rate enhancements.
    \item Deep JWST surveys such as COSMOS-Web are unlikely to serendipitously discover large numbers of TDEs in single-epoch observations due to their limited areas. However, they are sensitive to very high-redshift ($z>3$) TDEs. Multi-epoch monitoring of deep JWST fields, when combined with overlapping survey footprints, provides a path toward discovering the highest-redshift TDEs and probing the seeds of the first SMBHs.
\end{itemize}

By leveraging the complementary observations of LSST, Roman, and JWST, future TDE samples will enable population-level measurements of the evolution of the TDE rate, and the SMBH mass function. This work provides a framework for interpreting those measurements and shows that TDE samples in upcoming surveys will play a central role in constraining the origin and growth of supermassive black holes.

\section{Acknowledgments}

We thank the Transient Science at Space Telescope (TSST) group and the COSMOS-Web collaboration for insights into high-redshift transients and the high-redshift Universe. We also thank Massimo Stiavelli for useful discussions during the genesis of this work. This material is based upon work supported by the National Science Foundation Graduate Research Fellowship under Grant No. DGE2139757. We thank the director of Space Telescope Science Institute for the discretionary program DD: 3956.

\clearpage

\bibliographystyle{aasjournalv7}
\bibliography{rates}

\end{document}